\documentclass[aps,prb,reprint,superscriptaddress,floatfix,amsfonts,amsmath]{revtex4-1}
\usepackage{array}
\usepackage{amssymb}
\usepackage{amsmath}
\usepackage{graphicx}
\usepackage{contour}
\usepackage[bookmarks=false,colorlinks=true,urlcolor=blue,citecolor=blue,linkcolor=blue]{hyperref}
\usepackage{tikz}
\usetikzlibrary{calc,arrows}

\usepackage[compat=1.1.0]{tikz-feynman}
\tikzfeynmanset{/tikzfeynman/warn luatex=false}

\usepackage[version=4]{mhchem}

\newcommand{\mycomment}[1]{}
\newcommand{\note}[1]{#1}
\newcommand{\markup}[1]{#1}

\newcommand{\be}{\begin{equation}}
\newcommand{\ee}{\end{equation}}
\newcommand{\bea}{\begin{eqnarray}}
\newcommand{\eea}{\end{eqnarray}}
\newcommand{\beal}{\begin{align}}
\newcommand{\eeal}{\end{align}}
\newcommand{\bes}{\begin{equation} \begin{split}}
\newcommand{\ees}{\end{split} \end{equation}}

\newcommand{\f}{\frac}

\newcommand{\tr}{\mbox{Tr}}

\newcommand{\order}[1]{{\cal O}(#1)}

\newcommand{\qv}{\vec{q}}
\newcommand{\qvp}{\vec{q}^{\,\prime}}

\newcommand{\Qv}{\vec{Q}}
\newcommand{\pv}{\vec{p}}

\newcommand{\rv}{\vec{r}}

\newcommand{\Sv}{\vec{S}}

\newcommand{\Pimat}{\mathbf{\Pi}}
\newcommand{\Sigmamat}{\mathbf{\Sigma}}
\newcommand{\Dmat}{\mathbf{D}}
\newcommand{\Dinvmat}{\mathbf{D}^{-1}}
\newcommand{\Dzeromat}{\mathbf{D_0}}
\newcommand{\Dzeroinvmat}{\mathbf{D^{-1}_0}}
\newcommand{\Lambdamat}{\mathbf{\Lambda}}

\newcommand{\Kmat}{\mathbf{K}}
\newcommand{\Keffmat}{\mathbf{K_{\mathrm{eff}}}}
\newcommand{\Keffinvmat}{\mathbf{K}^{-1}_{\mathbf{\mathrm{eff}}}}
\newcommand{\Kinvmat}{\mathbf{K}^{-1}}
\newcommand{\Kinv}{\ensuremath{K^{-1}}}

\newcommand{\Keffinv}{K^{-1}_{\mathrm{eff}}{}}

\newcommand{\xhat}{\hat{x}}
\newcommand{\yhat}{\hat{y}}

\newcommand{\Jonethree}{$J_1$-$J_3$}

\newcommand{\Jonetwothree}{$J_1$-$J_2$-$J_3$}
\newcommand{\Srest}{\ensuremath{S_{R}}}
\newcommand{\singleq}{single-$\vec{q}$ }

\newcommand{\RN}[1]{%
\textup{\uppercase\expandafter{\romannumeral#1}}%
}

\newcommand{\abs}[1]{\lvert #1 \rvert}

\definecolor{taylorswift}{rgb}{0.0862745098,0.4666666667,0.3411764706}
\definecolor{fearless}{rgb}{0.8862745098,0.6117647059,0.2823529412}
\definecolor{speaknow}{rgb}{0.4588235294,0.2274509804,0.4980392157}
\definecolor{red}{rgb}{0.6509803922,0.1254901961,0.2705882353}
\definecolor{TS1989}{rgb}{0.1803921569,0.6,0.9764705882}
\definecolor{reputation}{rgb}{0.1450980392,0.1490196078,0.1529411765}
\definecolor{lover}{rgb}{0.8392156863,0.2117647059,0.5529411765}

\begin{document}



\title{\markup{Arc-shaped structure factor} in the {\Jonetwothree} classical Heisenberg model on the triangular lattice}


\author{Cecilie Glittum}
\affiliation{Department of Physics, University of Oslo, P.~O.~Box 1048 Blindern, N-0316 Oslo, Norway}

\author{Olav  F.~\surname{Sylju{\aa}sen}}
\affiliation{Department of Physics, University of Oslo, P.~O.~Box 1048 Blindern, N-0316 Oslo, Norway}

\date{\today}


\begin{abstract}
We study the {\Jonetwothree} classical Heisenberg model with ferromagnetic $J_1$ on the triangular lattice using the Nematic Bond Theory. For parameters where the momentum space coupling function $J_{\qv}$ shows a discrete set of minima, we find that the system in general exhibits a single first-order phase transition between the high-temperature ring liquid and the low-temperature single-$\qv$ planar spiral state. Close to where $J_{\qv}$ shows a continuous minimum, we on the other hand find several phase transitions upon lowering the temperature. Most interestingly, we find an intermediate temperature \markup{``arc" regime}, where the structure factor breaks rotational symmetry and shows a broad \markup{arc-shaped} maximum. We map out the parameter region over which this \markup{arc regime} exists and characterize details of its static structure factor over the same region.
\end{abstract}

\maketitle

\section{Introduction \label{sec:introduction}}
\noindent
The Mermin-Wagner theorem\cite{MerminWagner1966} forbids magnetic long-range order in two-dimensional Heisenberg magnets at finite temperatures. Nevertheless, such magnets may still exhibit phase transitions where a {\em discrete} point group symmetry of the lattice is broken. The type of order to expect in such cases
is usually that of a \singleq planar spiral state with a pitch vector taken from the set of wave vectors $\Qv$ that minimize the coupling function in momentum space $J_{\qv}$. Lattice point group symmetries will transfer the $\Qv$s into one another, and can be broken if the different $\Qv$s correspond to inequivalent spin states under global continuous spin rotations.\cite{Villain1977}

This scenario becomes more complicated when the $\Qv$s form a continuous set. In those cases the entropy, in contrast to the energy $J_{\qv}$, may favor a discrete subset of the $\Qv$s and so there can still be phase transitions breaking lattice point group symmetries at finite temperatures. This \markup{{\em order by disorder}} scenario\cite{Villain1980,Henley1989,CCL1990} happens in particular for the Heisenberg antiferromagnet on the honeycomb lattice for sufficiently large second neighbor coupling,\cite{Mulder2010,Okumura2010} and on the square lattice when a third neighbor coupling is included.\cite{Seabra2016} In all these cases, the order to expect can be inferred by finding the $\Qv$s corresponding to maximal spin wave entropy.

Here we investigate the lattice symmetry breaking phase transitions of the classical Heisenberg
model on the triangular lattice. Spontaneous breaking of lattice symmetries does not happen for the nearest neighbor model. Therefore, we add second and third neighbor interactions \markup{as shown in Fig.~\ref{fig:lattice}. The Hamiltonian is

\begin{equation}
H = J_1 \sum_{\langle i,j \rangle} \vec{S}_i\cdot \vec{S}_j + J_2 \! \sum_{\langle\langle i,j \rangle\rangle} \! \vec{S}_i\cdot \vec{S}_j + J_3 \!\!\! \sum_{\langle\langle\langle i,j \rangle\rangle\rangle} \!\!\! \vec{S}_i\cdot \vec{S}_j.
\end{equation}

\begin{figure}[h]
\includegraphics[width=0.28\textwidth]{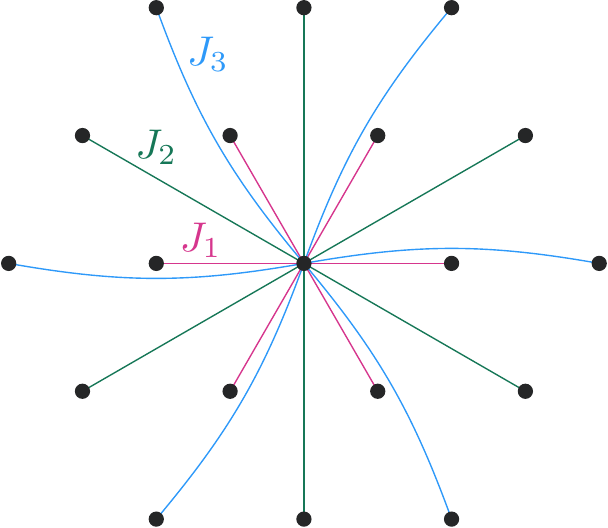}
\caption{The triangular lattice with up to third nearest neighbor interactions.\label{fig:lattice}}
\end{figure}}

\markup{This {\Jonetwothree} Heisenberg model has several distinct phases at zero temperature.\cite{Rastelli1979} At finite temperatures in a magnetic field it is known to have a Skyrmion lattice phase.\cite{Okubo2012} It has been proposed as a model for \ce{NiGa2S4},\cite{Nakatsuji2007,Mazin2007,Tamura2008} and its spin-1/2 version has been studied in the context of quantum spin liquids.\cite{Iaconis2018,Gong2019}} \markup{The extended couplings} allow us to tune $J_{\qv}$ between discrete and continuous minima.
For this {\Jonetwothree} Heisenberg model with $J_1$ ferromagnetic, we also find the \markup{order by disorder} scenario, but it plays out in an interesting way. Our main result is that the ordering occurs via a sequence of {\em two} phase transitions as the temperature is lowered. \markup{Particularly interesting is the intermediate phase, where the static structure factor is dominated by an arc-shaped ridge. This arc breaks lattice rotational symmetry, but not all mirror symmetries, and is not a single-$\qv$ state.}

To be able to efficiently investigate large portions of parameter space, we employ the Nematic Bond Theory (NBT),\cite{Schecter2017} which is a set of approximate self-consistent equations for classical Heisenberg magnets. The equations can be solved numerically for large lattices.\cite{Syljuasen2019} Besides calculating order parameters and correlation functions, we show here that the NBT can also be used to calculate the free energy directly, which allows us to determine the order of the phase transitions. We explain the NBT with an emphasis on how to obtain the free energy in section~\ref{sec:method}. The details of the {\Jonetwothree} model on the triangular lattice are given in section~\ref{sec:Jonetwothree-model}, and the results are presented in section \ref{sec:results}. We end with a discussion in section~\ref{sec:discussion}.

\section{Method \label{sec:method}}
\noindent
The NBT is conveniently formulated in momentum space:
\be
H = \sum_{\qv} J_{\qv} \Sv_{-\qv} \cdot \Sv_{\qv},
\label{qHamiltonian}
\ee
where the sum goes over the first Brillouin zone.
\mycomment{
\\
Fourier conventions:
\be
f_{\rv} = \f{1}{V^{\alpha}} \sum_{\qv} f_{\qv} e^{i \qv \cdot \rv}, \qquad
f_{\qv} = \f{1}{V^{1-\alpha}} \sum_{\rv} f_{\rv} e^{-i \qv \cdot \rv}, \nonumber
\ee
where $\alpha$ is replaced by the appropriate greek letter: $\mu=1/2$ for spins, $\gamma=1$ for $J$ (for $J$ there is also an extra factor 2 so that
$J_{\qv} = \f{1}{2} \sum_{\rv} J_{\rv} e^{-i \qv \cdot \rv}$.), $\nu=\nu^\prime=0$ for $\lambda$.}

The classical spins on all sites are unit length vectors: $|\Sv_{\rv}|=1$. These length constraints are enforced in the partition function as integral representations of $\delta$-functions
\be
\delta \left( |\Sv_{\rv}| -1 \right) = \int_{-\infty}^\infty \! \f{\beta d \lambda_{\rv}}{\pi} \; e^{-i \beta \lambda_{\rv} \left( \Sv_{\rv} \cdot \Sv_{\rv} - 1 \right)},
\ee
where we have scaled the integration variable $\lambda_{\rv}$ by the inverse temperature, $\beta=1/T$. This gives the partition function
\be
Z = \int \! \! D\Sv \, d\Delta \, D\lambda \, e^{-\beta\sum_{\qv,\qvp}\left( \Kmat_{\qv,\qvp} - \Lambdamat_{\qv,\qvp} \right) \Sv_{\qv}^* \cdot \Sv_{\qvp}  + \beta V \Delta},
\ee
where we have introduced a momentum space matrix $\Lambdamat_{\qv,\qvp} \equiv -i \lambda_{\qv-\qvp} (1-\delta_{\qv,\qvp})$, and $\lambda_{\qv}$ is the Fourier-transformed constraint integration variable. We have separated out its $\qv =0$ component and written it as  $\Delta \equiv i \lambda_{\qv=0}$ and put it into another momentum space matrix $\Kmat_{\qv,\qvp} \equiv K_{\qv} \; \delta_{\qv,\qvp}$, where $K_{\qv} \equiv J_{\qv} + \Delta$. The integration measures are always redefined to include factors of volume $V$, $\beta$, $\pi$ and $-i$.  \markup{The inverse of $K_{\qv}$ is essentially the spin-spin correlation function in momentum space, and $\Delta$ can be interpreted as the average constraint, similar to the self-consistent field in the self-consistent Gaussian approximation.  The NBT goes beyond this as it also accounts for the fluctuations $\Lambdamat_{\qv,\qvp}$ around the average constraint. This is essential in order to capture lattice point group symmetry breaking phase transitions.}

\mycomment{
\be
D \Sv \, d\Delta \, D\lambda \equiv \Pi_{\qv} d \Sv_{\qv} \left(\f{\beta V^{1/2}}{\pi} (-i)d \Delta \right) \Pi_{\qv \neq 0} \f{\beta V^{1/2}}{\pi} d{\lambda_{\qv}}, \nonumber
\ee
where on the right hand side $d \Delta = i d \lambda_{\qv=0}$. The factors $V^{1/2}$ comes from the Jacobian associated with going from $\lambda_{\rv}$ to $\lambda_{\qv}$ with the definition $\lambda_{\rv} = \sum_{\qv} \lambda_{\qv} e^{i \qv \cdot \rv}$.
}

The integrals over the spin components can now be taken as independent Gaussian integrals. We generalize the spins to have $N_s$ vector components, but will set $N_s=3$ at the end of the calculation. We scale the spin components by a factor $1/\sqrt{\beta}$ and perform the Gaussian integrals to get
\be
Z = \int \! \! d\Delta \, D\lambda \, e^{-S[\Delta,\lambda]},
\ee
where the effective constraint action is
\be
S[\Delta,\lambda] \equiv \f{N_s}{2}  \tr  \ln{\left( \Kmat - \Lambdamat \right)} - \beta V \Delta.
\ee
\mycomment{
\\
We have redefined the integration measure with appropriate factors of $\beta$ and $\pi$.
The new integration measure is
\be
d \Delta D \lambda \equiv \left( \f{\beta V^{1/2}}{\pi} \right)^V \left( \f{\pi}{\beta} \right)^{N_s V/2}
(-i) d \Delta \Pi_{\qv \neq 0} d \lambda_{\qv} \nonumber
\ee
}
Expanding this expression in powers of $\Lambdamat$, we get
\be
S[\Delta,\lambda] = -\beta V \Delta + \f{N_s}{2}  \tr  \ln{\Kmat}
+ \f{1}{2} \sum_{\qv \neq 0} \lambda_{-\qv} D_{0,\qv}^{-1} \lambda_{\qv} + S_r,
\ee
where we have used the quadratic term in $\Lambdamat$ to give the inverse constraint propagator $\Dzeroinvmat_{\! \! \qv \qvp} \equiv D^{-1}_{0,\qv} \; \delta_{\qv,\qvp}$ with
\be \label{eq:D0qinv}
D_{0,\qv}^{-1} = \f{N_s}{2} \sum_{\pv} \Kinv_{\pv+\qv} \Kinv_{\pv},
\ee
and the interaction $S_r$ is
\be
S_r = -\f{N_s}{2} \sum_{n=3}^{\infty} \f{1}{n} \tr \left( \Kinvmat \Lambdamat \right)^n.
\ee
There is no linear term in $\Lambdamat$ because $\Lambdamat$ has no diagonal components, which follows from separating out $\lambda_{\qv=0}$.

We then treat $S_r$ as a perturbation about the Gaussian action defined by the quadratic terms in $\lambda$ and integrate over $\lambda$ so that
\be
Z = \int d \Delta e^{-S[\Delta]},
\ee
where
\be
S[\Delta] \equiv -\beta V \Delta + \f{N_s}{2} \tr \ln{\Kmat} + \f{1}{2} \tr \ln{\Dzeroinvmat} - \ln{ \langle e^{-S_r} \rangle}.
\ee
The brackets $\langle \rangle$ indicate an average with respect to the Gaussian action.
\mycomment{
\\
The new measure reads
\be
d \Delta \equiv \left( \f{\beta V^{1/2}}{\pi} \right)^V \left( \f{\pi}{\beta} \right)^{N_s V/2}
\left( 2\pi \right)^{(V-1)/2}
(-i) d \Delta \nonumber
\ee
}

\begin{figure}[t]
\includegraphics[width=0.1\textwidth]{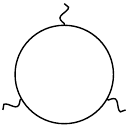}
\caption{A ring with 3 wavy hooks; the $n=3$ term in $S_r$.\label{ringdiagrams}}
\end{figure}

\begin{figure}[t]
\includegraphics[width=0.4\textwidth]{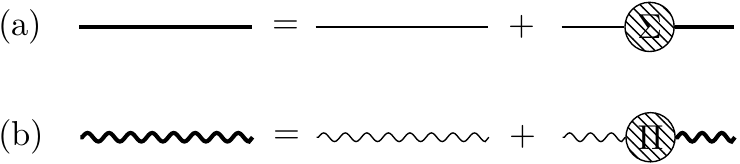}
\caption{Dyson equations for (a) the renormalized spin propagator $\Keffinv$ (bold solid line), and (b) the renormalized constraint propagator $D$ (bold wavy line). \label{Dyson}}
\end{figure}

The perturbation theory can be formulated diagrammatically with solid and wavy lines indicating $\Kinv$ and $D_0$ respectively. Interactions in $S_r$ are ring diagrams having hooks \markup{where} wavy lines can attach, see Fig.~\ref{ringdiagrams}.
%
%
We then use a self-consistent procedure where a self-energy $\Sigmamat_{\qv,\qvp} \equiv \Sigma_{\qv}\,\delta_{\qv,\qvp}$ and a polarization $\Pimat_{\qv,\qvp} \equiv \Pi_{\qv}\,\delta_{\qv,\qvp}$ are defined to renormalize $\Kinvmat$ and $\Dzeromat$ respectively according to the Dyson equations shown in Fig.~\ref{Dyson}.

\begin{figure}[t]
\includegraphics[width=0.35\textwidth]{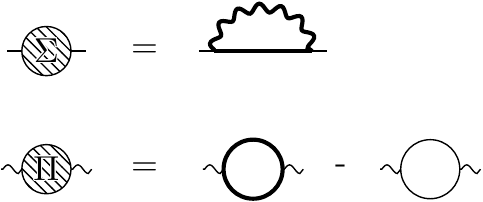}
\caption{Self-consistent equations for the self-energy and the polarization. The bold lines on the right hand sides also include the self-energy and the polarization. \label{selfconsistentdiagrams}}
\end{figure}

\begin{figure}[t]
\includegraphics[width=0.4\textwidth]{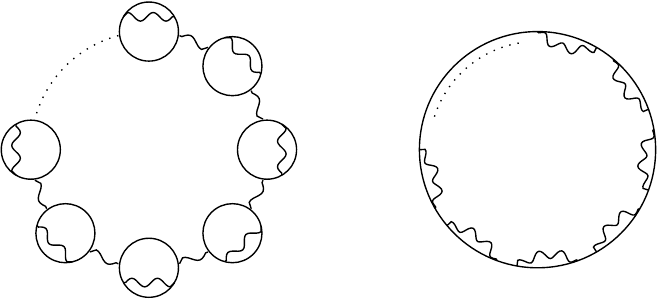}
\caption{Diagrams included in the free energy. \label{includeddiagrams}}
\end{figure}

\begin{figure}[t]
\includegraphics[width=0.30\textwidth]{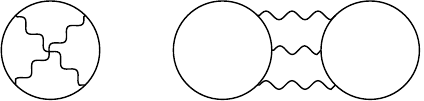}
\caption{Leading order non-vanishing diagrams in \Srest. \label{vertexcorrectiondiagrams}}
\end{figure}

The Dyson equations yield $\Keffmat = \Kmat - \Sigmamat$ and $\Dinvmat = \Dmat_{0}^{-1} - \Pimat$. The self-energy and the polarization are next approximated self-consistently by the diagrams in Fig.~\ref{selfconsistentdiagrams}, which are equivalent to the equations
\begin{align}
\Sigma_{\qv} &= - \sum_{\pv \neq 0} \Keffinv_{\qv-\pv} D_{\pv}, \label{selfenergy_eq} \\
\Pi_{\qv} &= - \f{N_s}{2} \sum_{\pv} \Keffinv_{\pv+\qv} \Keffinv_{\pv} + \f{N_s}{2} \sum_{\pv} \Kinv_{\pv+\qv} \Kinv_{\pv}.
\label{polarization_eq}
\end{align}
Combining the Dyson equation for $\Dinvmat$ with Eqs.~\eqref{eq:D0qinv} and~\eqref{polarization_eq}, the renormalized constraint propagator becomes
\be
D_{\qv}^{-1} = \f{N_s}{2} \sum_{\pv} \Keffinv_{\pv+\qv} \Keffinv_{\pv}. \label{constraintprop_eq}
\ee
The unrenormalized propagators can be expressed in terms of their renormalized equivalents so that $S[\Delta]$ becomes
\begin{align}
S[\Delta] &= -\beta V \Delta + \f{N_s}{2} \tr \ln{\Keffmat} + \f{1}{2} \tr \ln{\Dinvmat} \nonumber \\
& \quad +\f{N_s}{2} \tr \left( \Keffinvmat \Sigmamat \right) + \Srest,  \label{SofDelta}
\end{align}
where the remainder \Srest\ is defined in appendix \ref{app:Sremainder}.
In the following we will simply omit $\Srest$, which means that after this omission $S[\Delta]$ includes all diagrams of the sort shown in Fig.~\ref{includeddiagrams}, but neglects, among others, diagrams with vertex corrections shown in Fig.~\ref{vertexcorrectiondiagrams}.

The final integral over $\Delta$ is performed using the saddle point approximation, see appendix \ref{app:Saddlepoint}, which gives the condition
\be
\f{N_s T}{2V} \sum_q \Keffinv_{\qv} = 1 \label{saddlepoint}.
\ee
By taking also into account the Gaussian fluctuations in $\Delta$ about the saddle point value and restoring omitted constants,
\mycomment{
\\
The Gaussian fluctuations about the saddle point gives an extra constant factor $\sqrt{2\pi}$. Thus pulling out all the constant factors in front of the integral gives
\begin{align}
Z &=
\left( \f{\beta V^{1/2}}{\pi} \right)^V \left( \f{\pi}{\beta} \right)^{N_s V/2}
\left( 2\pi \right)^{V/2} \int_{-\infty}^\infty d \lambda_{\qv=0} e^{-S[i \lambda_{\qv=0}]}
\nonumber \\
& =  (2V)^{V/2} \pi^{(N_s-1)V/2} \beta^V \beta^{-N_sV/2} \int_{-\infty}^\infty d \lambda_{\qv=0} e^{-S[i \lambda_{\qv=0}]} \nonumber \\
& =  e^{ \f{V}{2} \ln{(2V)} + \f{(N_s-1)V}{2} \ln{\pi} + V \ln{\beta} -\f{N_s V}{2} \ln{\beta}} \int_{-\infty}^\infty d \lambda_{\qv=0} e^{-S[i \lambda_{\qv=0}]} \nonumber \\
& =  e^{-\beta V \left(  -\f{1}{2\beta} \ln{(2V)} - \f{(N_s-1)}{2\beta} \ln{\pi} - \f{1}{2\beta} \ln{(\beta^2)} + \f{N_s}{2\beta} \ln{\beta}  + \f{1}{\beta V} S[\Delta_0] + \f{1}{2\beta V} \ln D^{-1}_{\qv=0} \right)} \nonumber
\end{align}
}
we find the following expression for the free energy density $f = -\f{1}{\beta V} \ln Z$:
\begin{align}
f &=  - \Delta - \f{N_s T}{2V} \sum_{\qv} \ln{\left( T \Keffinv_{\qv}\right)} +\f{N_s T}{2V} \sum_{\qv} \Keffinv_{\qv} \Sigma_{\qv} \nonumber \\
& \quad + \f{T}{2V} \sum_{\qv} \ln{\left(T^2 D^{-1}_{\qv}/2V \right)} -\f{(N_s-1)T}{2} \ln{\pi},  \label{freeenergydensity}
\end{align}
where the $\ln{D_{\qv}^{-1}}$-- sum also includes the $\qv=0$ term. This expression is similar to that used in Ref.~\onlinecite{Barci2013} in the context of the self-consistent screening approximation.
\mycomment{
\\
Note that in the master note we have defined
\begin{align}
\note{\Keffinv_{\qv}} &= \f{T}{J_{\qv} + \Delta - \Sigma_{\qv}} \nonumber \\
\note{D^{-1}_{\qv}} &= \f{N_s}{2} \sum_{\pv}
\f{T^2}{\left( J_{\qv+\pv} + \Delta - \Sigma_{\qv+\pv} \right) \left( J_{\pv} + \Delta - \Sigma_{\pv} \right)} \nonumber \\
\note{\Sigma_{\qv}} &= - \sum_{\pv \neq 0}  \note{\Keffinv_{\qv-\pv}} \note{D_{\pv}} \nonumber
\end{align}
which means that $\note{\Sigma}$ in the master note is $\Sigma$ here divided by $T$, meaning that $\Keffinv \Sigma$ is the same expression in the master note and here.

The inverse volume factor in the $\ln{(D^{-1})}$--term is correct. It makes the argument of the log independent of $V$. With the chosen Fourier convention of the $\lambda$'s; $D^{-1}$ is defined as a q-sum over a $V$-independent quantity, with no $1/V$ factor in front of the sum. The sum will therefore increase with $V$. The $1/V$ in the log which comes from the integration measure and the very same Fourier convention fixes this.

Note that in the computer code $\Sigma$ is defined with the opposite sign, so that it is always positive.
}

We solve the self-consistent equations \eqref{selfenergy_eq} and \eqref{constraintprop_eq} numerically, as described in details in Ref.~\onlinecite{Syljuasen2019}, and obtain expressions for $\Keffinv$, $D$ and $\Sigma$, which are then used to compute the free energy density from Eq.~(\ref{freeenergydensity}), and the static structure factor
\be
\mathcal{S}(\qv) \equiv \langle \Sv_{-\qv} \cdot \Sv_{\qv} \rangle = \f{N_s T}{2} \Keffinv_{\qv},
\ee
as shown in Refs.~\onlinecite{Schecter2017, Syljuasen2019}. We note that the saddle point condition Eq.~(\ref{saddlepoint}) is equivalent to the condition $\langle \Sv_{\rv} \cdot \Sv_{\rv} \rangle =1$.

\section{{\Jonetwothree} model \label{sec:Jonetwothree-model}}
On the triangular lattice, the momentum space coupling function is
\begin{align}
J_{\qv} &= J_1 \left[ \cos{(q_1)} + \cos{(q_2)} + \cos{(q_3)} \right]
\nonumber \\
&+ J_2 \left[ \cos{( q_1 - q_2 )}
+ \cos{(q_2 - q_3)}
+ \cos{(q_3 - q_1)} \right] \nonumber \\
&+ J_3 \left[ \cos{(2 q_1)}+ \cos{(2q_2)}+ \cos{(2q_3)} \right], \label{jq}
\end{align}
where $q_i \equiv \qv \cdot \vec{a}_i$ and the lattice vectors are $\vec{a}_1=\xhat$, $\vec{a}_{2}=-\f{1}{2} \xhat+ \f{\sqrt{3}}{2} \yhat$ and
$\vec{a}_{3}=-\f{1}{2} \xhat - \f{\sqrt{3}}{2} \yhat$. The lattice spacing has been set to unity.
For further analysis, it is convenient to rewrite $J_{\qv}$ as
\be
J_{\qv} = 2J_3 \left[  \left( A_{\qv} - \f{1}{2} \left( 1 - \f{J_1}{2J_3} \right) \right)^2 \right]
+ \left( J_2 - 2J_3 \right) B_{\qv} +C,
\ee
where  $A_{\qv} \equiv \cos{(q_1)} + \cos{\left(q_2 \right)} + \cos{\left(q_3 \right)}$, $B_{\qv}~\equiv~\cos{\left(q_1-q_2 \right)} + \cos{\left(q_2-q_3 \right)} + \cos{\left( q_3-q_1 \right)}$,
and $C$ is a parameter-dependent constant.
\mycomment{ $C= -3J_3 -\f{J_3}{2} \left( 1 - \f{J_1}{2J_3} \right)^2$}
We will set $J_1=-1$ (FM) which defines our unit of energy.

\begin{figure}
\includegraphics[width=0.4\textwidth]{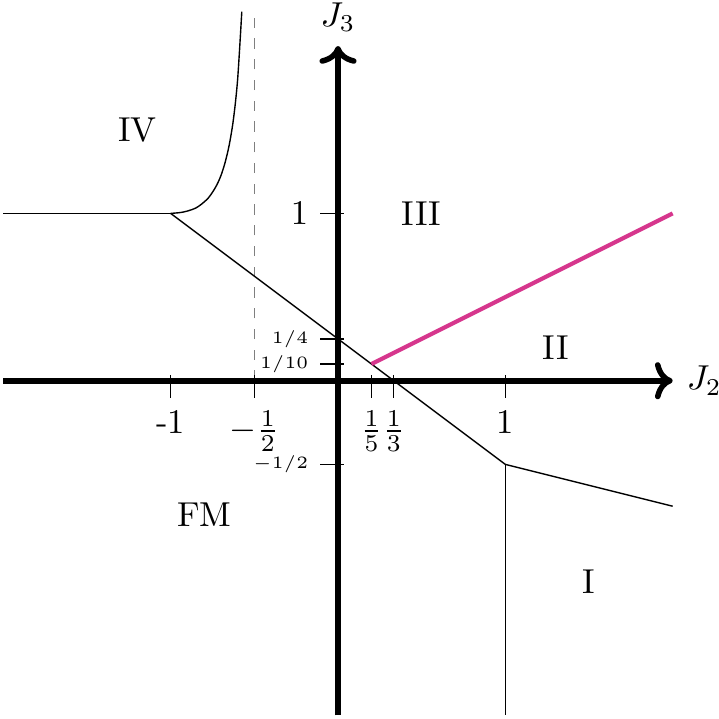}
\caption{Regions of different classes of wave vectors $\vec{Q}$ minimizing $J_{\qv}$ for ferromagnetic nearest neighbour coupling, \markup{$J_1=-1$}. The pink thick line shows the $\RN{2}$--$\RN{3}$ border, where the $\Qv$s form a continuous set.}
\label{fig: phase diagram ferro}
\end{figure}

\begin{figure}
\includegraphics[width=0.4\textwidth]{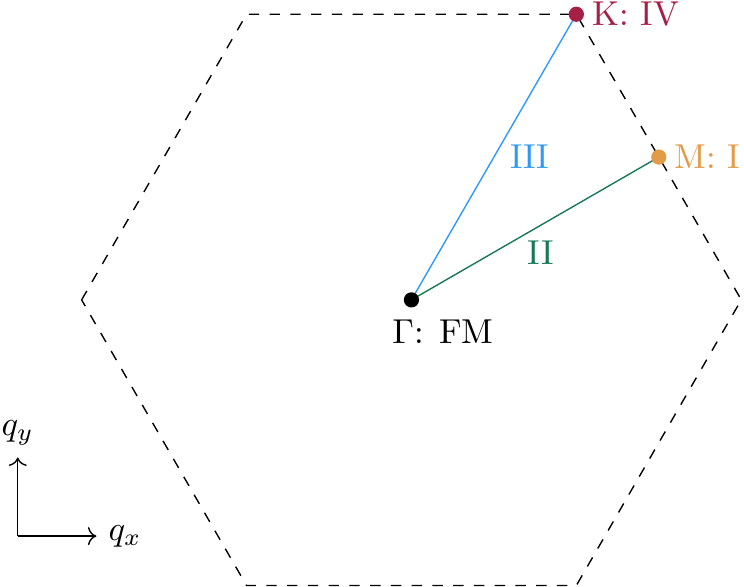}
\caption{Illustration of where the $\vec{Q}$s minimizing $J_{\qv}$ are located in reciprocal space for the different regions from Fig.~\ref{fig: phase diagram ferro}. The illustration is symmetric under rotations of $\frac{\pi}{3}$. The first Brillouin zone boundary is illustrated by the dashed lines. $\vec{Q}$ in the FM region is located at $\Gamma$ (black point). In regions $\RN{1}$ and $\RN{4}$ the $\vec{Q}$s are located at M (yellow points) and K (red points) respectively. Region $\RN{2}$ has $\vec{Q}$s along $\Gamma$M (green lines). In region $\RN{3}$, the $\vec{Q}$s lie on $\Gamma$K (blue lines).} 
\label{fig: Brillouin minimas}
\end{figure}

By minimizing $J_{\qv}$ with respect to $\qv$, we can find which single-$\qv$ states that minimize the energy. For generic choices of the parameters $J_2$ and $J_3$, these minimal $\Qv$s form a discrete set of symmetry-related points in the Brillouin zone. The different regions of $\Qv$s minimizing $J_{\qv}$ are shown in Fig.~\ref{fig: phase diagram ferro}, with the corresponding $\Qv$s illustrated in Fig.~\ref{fig: Brillouin minimas}. We define $\Gamma$M($\Gamma$K) as the lines connecting the $\Gamma$ point and the M(K) points, illustrated by the green(blue) lines in Fig.~\ref{fig: Brillouin minimas}.

As shown in Ref.~\onlinecite{Rastelli1979}, the length of the $\Qv$s minimizing $J_{\qv}$ in region $\RN{2}$ is given by
\begin{equation}\label{eq:QII}
Q_{\RN{2}} = \frac{2}{\sqrt{3}}\arccos\left(\frac{1 - J_2}{2J_2 + 4J_3}\right),
\end{equation}
while it in region $\RN{3}$ is given by
\begin{equation}\label{eq:QIII}
Q_{\RN{3}} = 2\arccos\left(\f{3J_2 - 2J_3 - \sqrt{(3J_2 + 2J_3)^2 + 8J_3}}{-8 J_3}\right).
\end{equation}

On the border between regions $\RN{2}$ and $\RN{3}$, where $J_2~=~2J_3$, the minimal $\Qv$s form a continuous set defined by
$A_{\Qv} = \f{1}{2} \left( 1 - \f{J_1}{J_2} \right)$. This collection of minimal $\Qv$s make a slightly deformed circular ring in momentum space. It is this border region which is of special interest in this paper.

\section{Results \label{sec:results}}

\subsection{Generic parameters}

\begin{figure}[t]
\begin{center}
\includegraphics[width=0.45\textwidth]{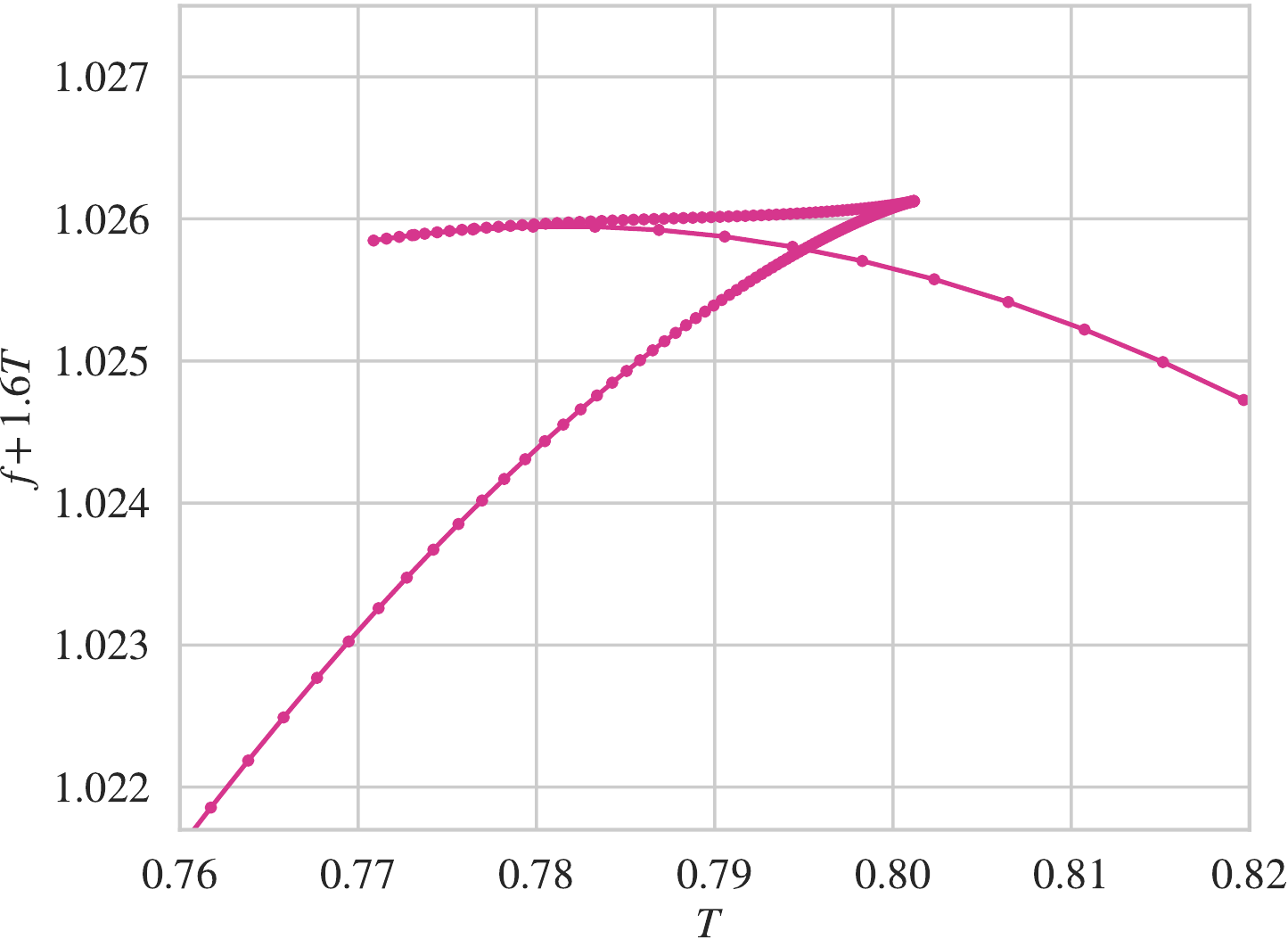}
\caption{Free energy density vs. $T$ for $(J_2,J_3) =(2,0)$. $L=200$. The free energy has been transformed by adding a linear term in $T$ in order to better visualize the discontinuity in its derivative. The free energy density is multivalued in the region $T \in [0.770,0.802]$.}
\label{fig:FreeE}
\end{center}
\end{figure}

\begin{figure}[t]
\begin{center}
\includegraphics[width=0.45\textwidth]{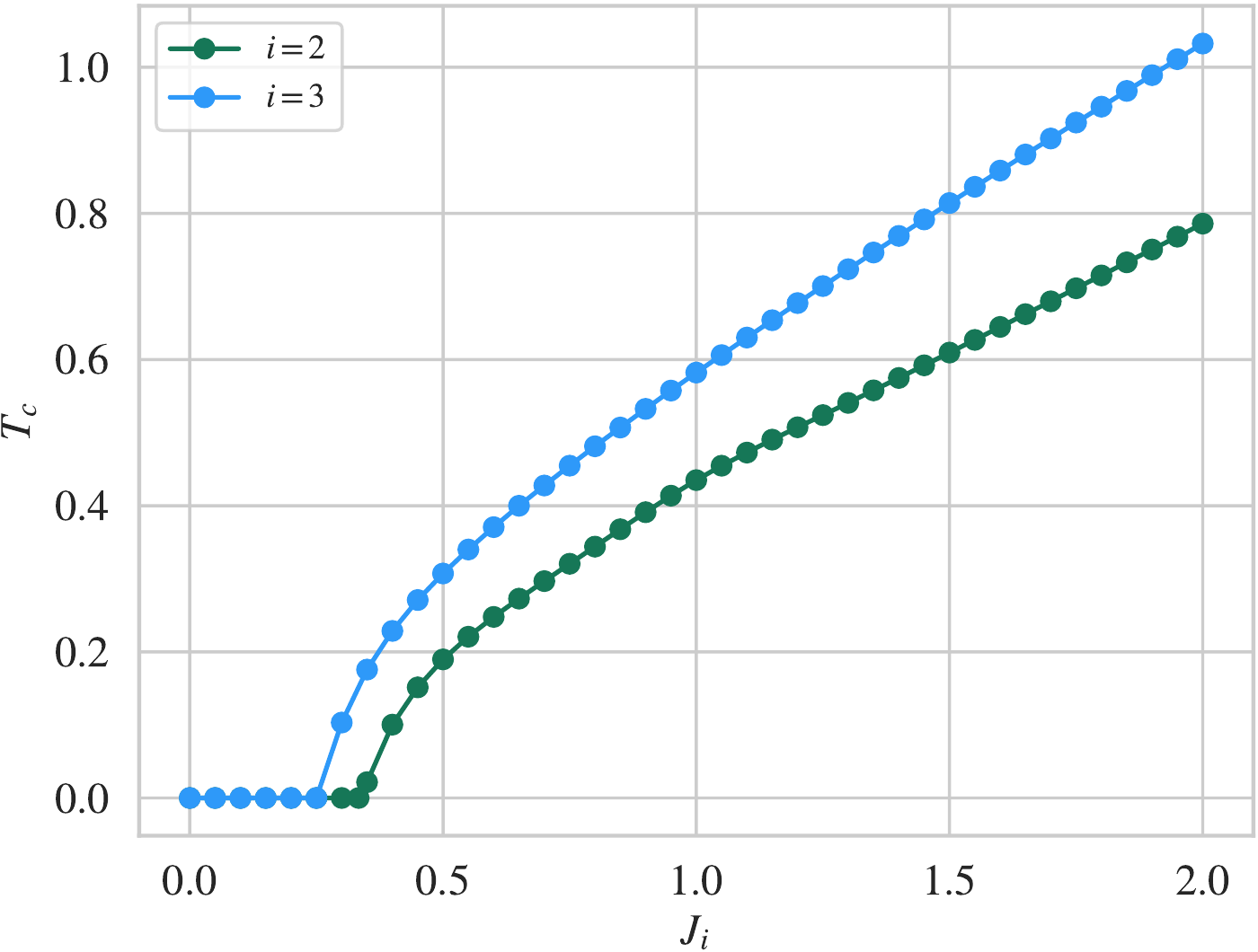}
\caption{Critical temperatures along the $J_2$- and $J_3$-axis.
\label{fig:axis_Tc}}
\end{center}
\end{figure}

For generic parameter values, the $\Qv$s form a discrete set, but it is only possible to break the point group symmetries of the lattice in regions $\RN{1}$, $\RN{2}$ and $\RN{3}$. Such symmetry breaking is not possible in region $\RN{4}$, as all configurations are equivalent by a global spin rotation. In the regions $\RN{1}$, $\RN{2}$ and $\RN{3}$, we in general find that the system exhibits a {\em single} first-order temperature-driven phase transition breaking rotational symmetry of the lattice.
In Fig.~\ref{fig:FreeE} we show as an example of this the free energy density as a function of $T$ for the point $(J_2,J_3)=(2,0)$ in region $\RN{2}$. From this figure, we see that there is a temperature-region where the free energy density is multivalued. \markup{This multivaluedness reflects the fact that there are multiple values of $\Delta$ with associated self-energies $\Sigma_q$ that lead to the same temperature when solving the saddle-point equation, Eq.~\eqref{saddlepoint}.} The thermodynamically stable states are those which minimize the free energy density. The existence of the corner point of the lowest free energy curve at $T_c= 0.795$ indicates a first-order phase transition there.
Repeating this for other parameter points $(J_2,0)$ and also for $(0,J_3)$, we find similar first-order phase transitions with critical temperatures given in Fig.~\ref{fig:axis_Tc}.

\markup{

\begin{figure}[t]
\begin{center}
\includegraphics[width=0.4\textwidth]{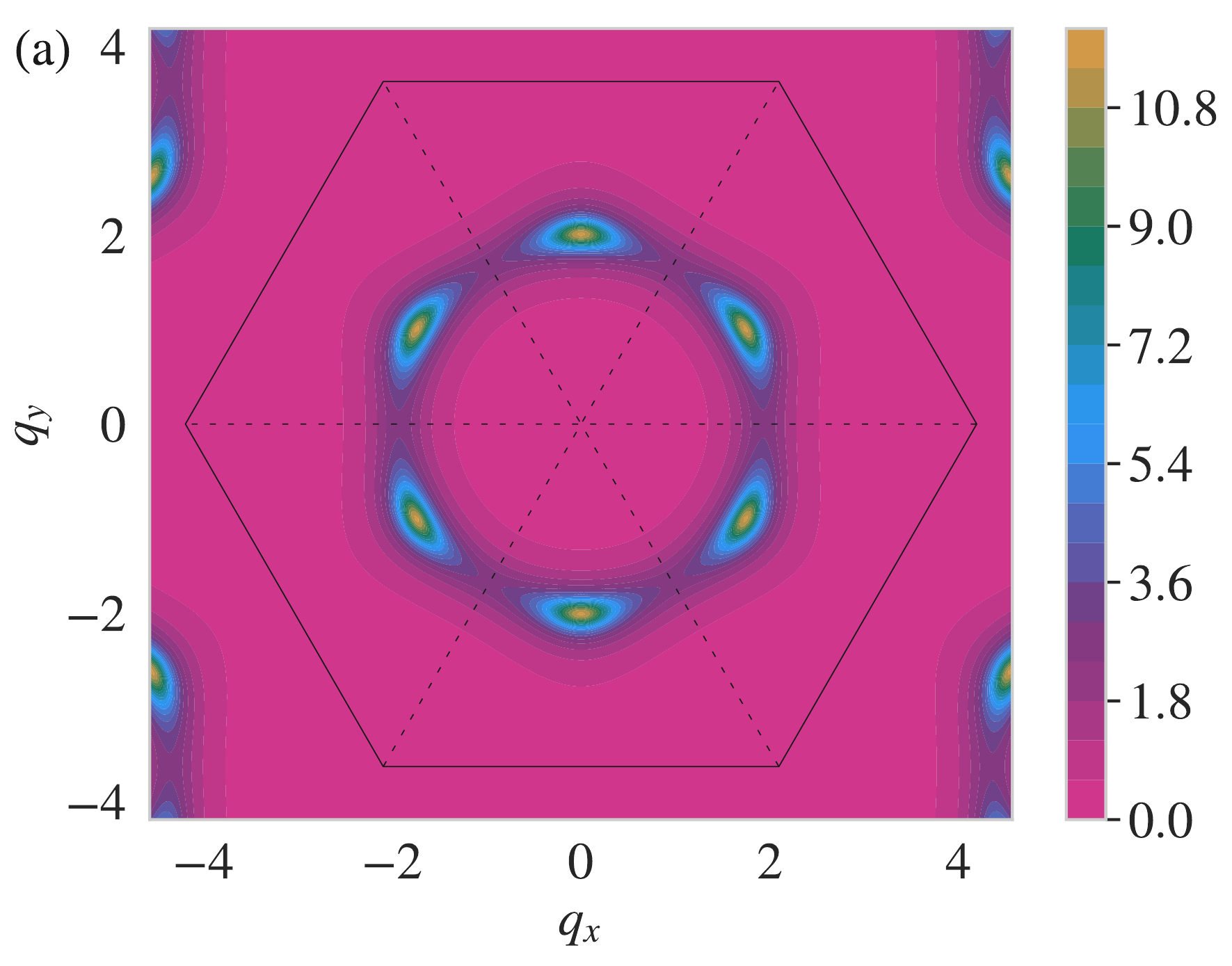}
\includegraphics[width=0.4\textwidth]{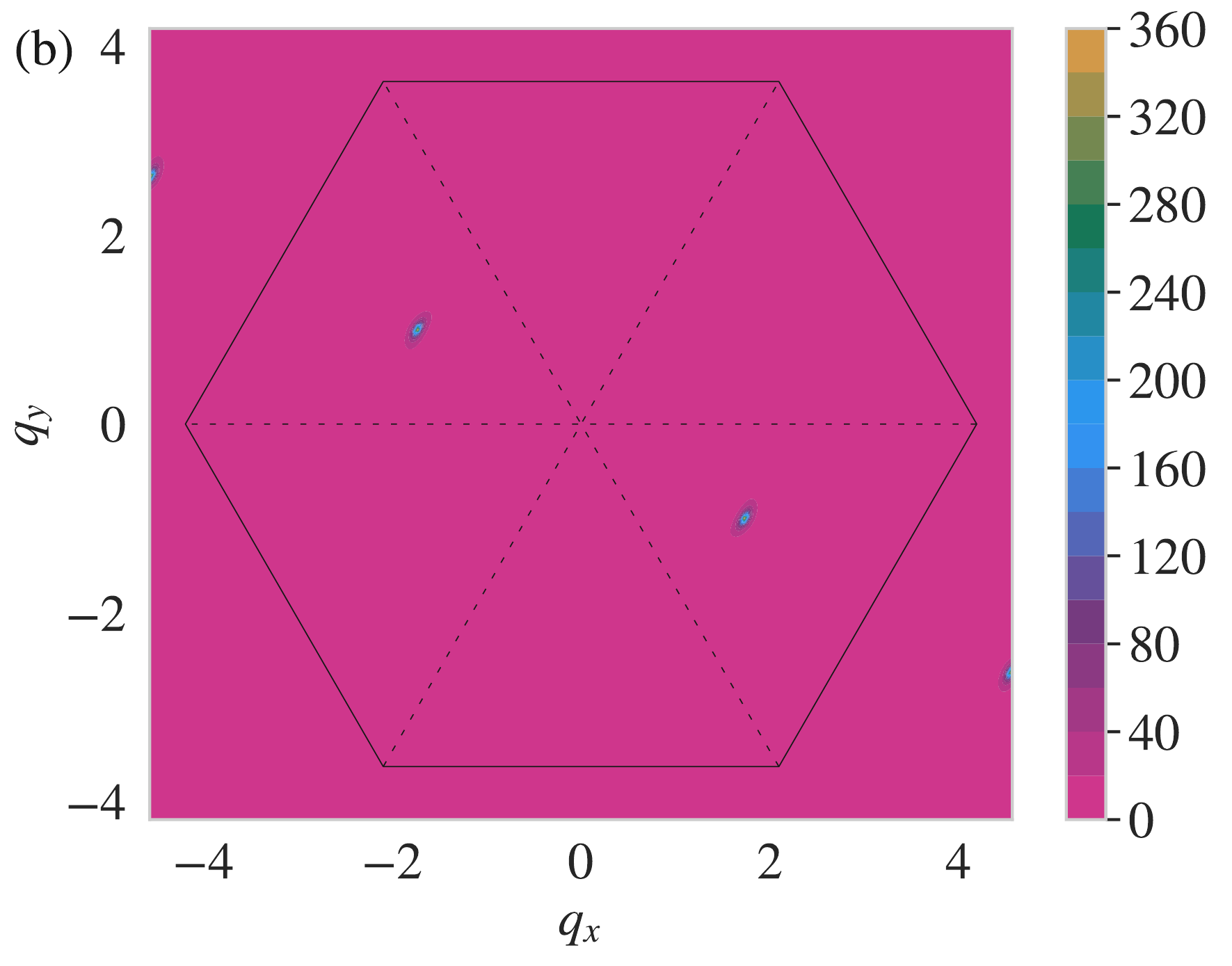}
\caption{Structure factors $\mathcal{S}(\qv)$ for a parameter point in region $\RN{2}$, $(J_2,J_3) = (2,0.5)$, for different temperatures: (a) high-$T$ ring liquid phase at $T=0.869$ and (b) low-$T$ single-$\qv$ phase with $\qv$ on $\Gamma$M at $T=0.868$. $L=200$.}
\label{fig:Keff_II}
\end{center}
\end{figure}

\begin{figure}[t]
\begin{center}
\includegraphics[width=0.4\textwidth]{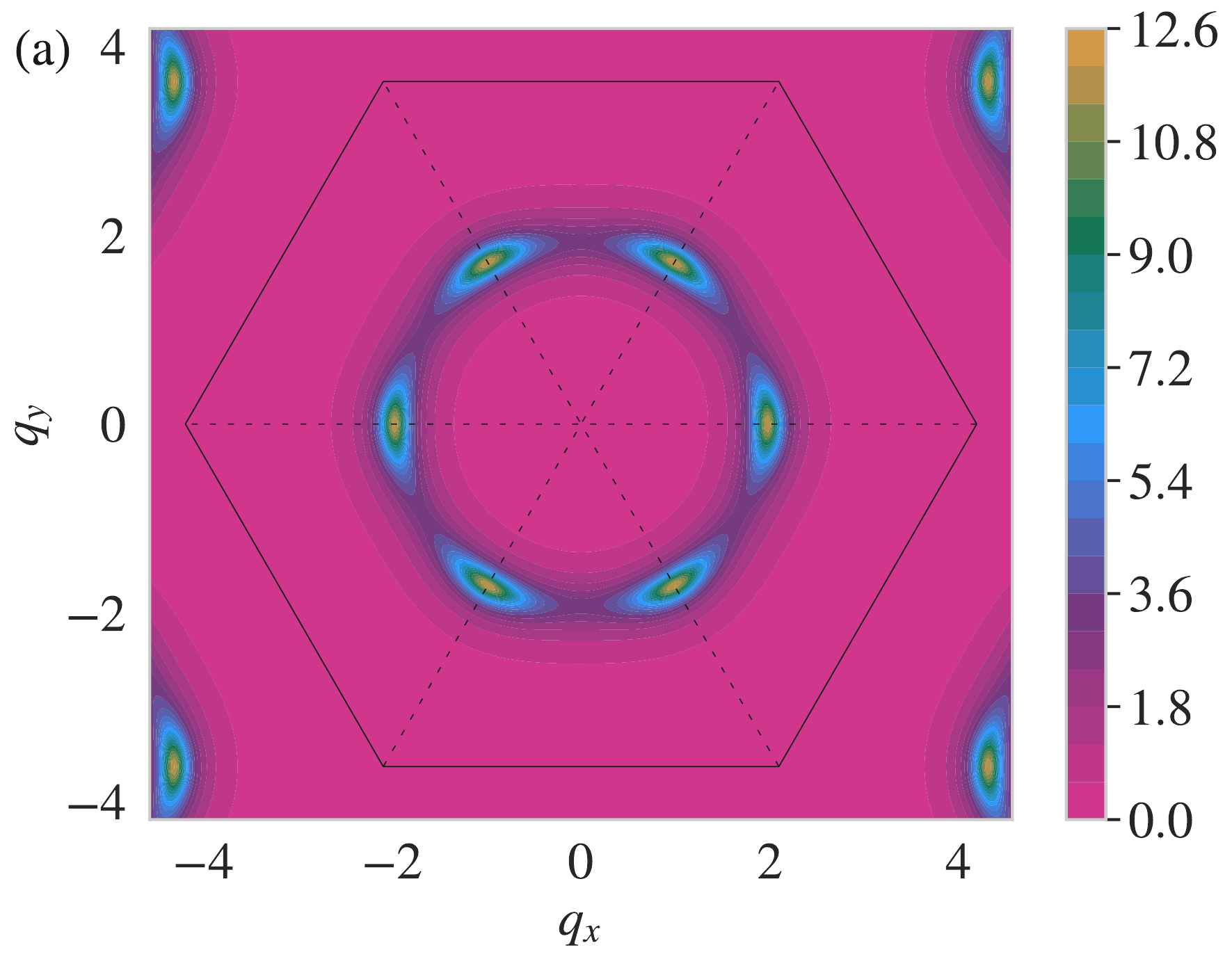}
\includegraphics[width=0.4\textwidth]{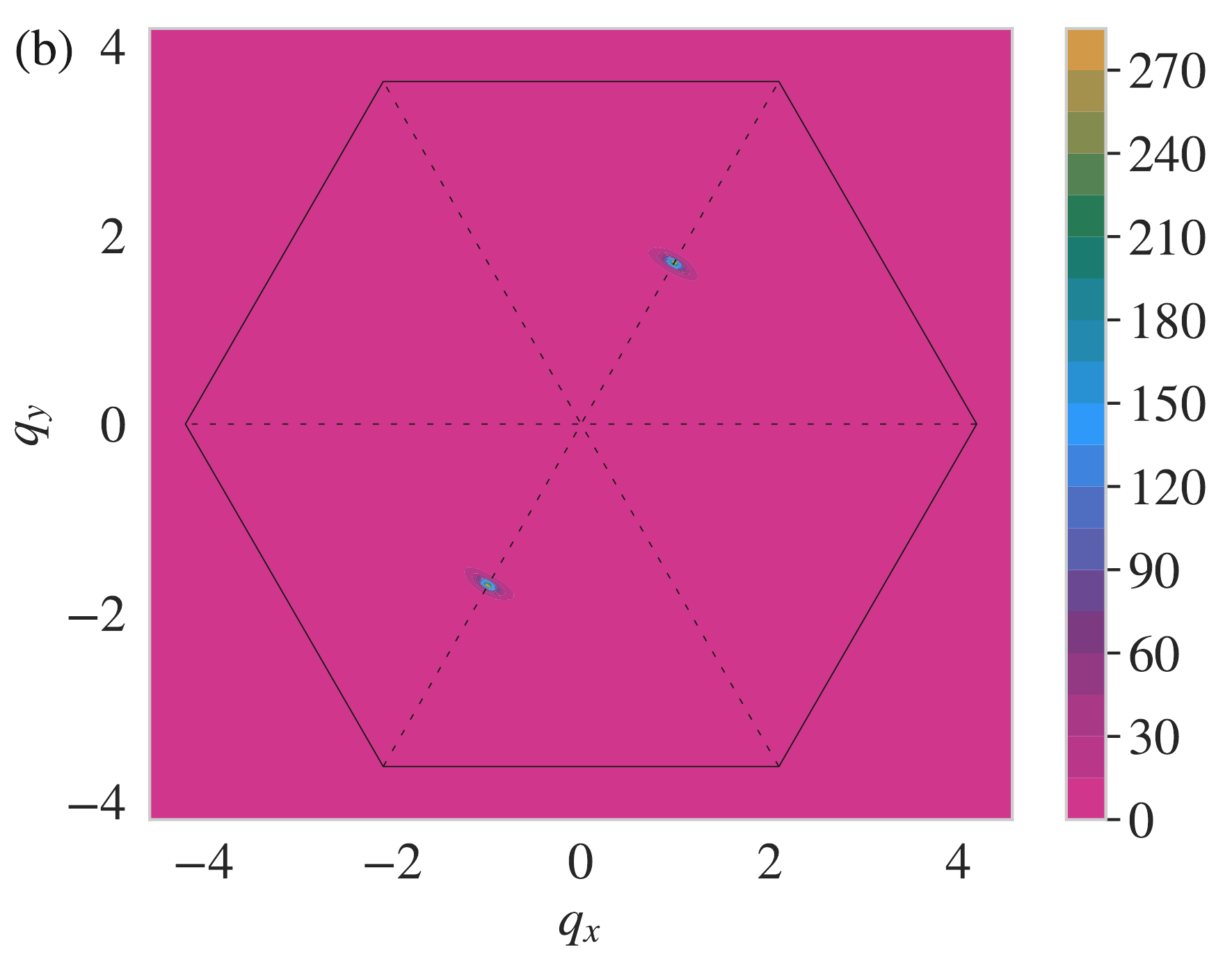}
\caption{Structure factors $\mathcal{S}(\qv)$ for a parameter point in region $\RN{3}$, $(J_2,J_3) = (2,1.5)$, for different temperatures: (a) high-$T$ ring liquid phase at $T=1.194$ and (b) low-$T$ single-$\qv$ phase with $\qv$ on $\Gamma$K at $T=1.191$. $L=200$.}
\label{fig:Keff_III}
\end{center}
\end{figure}

}

Such a phase transition is between a high-$T$ ring liquid phase where the static structure factor $\mathcal{S}(\qv)$ shows a ring-like
feature in momentum space and a low-$T$ phase where the system breaks the rotational symmetry of the lattice as it enters a single-$\qv$ spiral state, where the pitch vector is determined by one of the minimal $\Qv$s. Thus, in region $\RN{1}$ we generally get single-$\qv$ states with $\qv = $M and in region $\RN{2}$($\RN{3}$) we generally get single-$\qv$ states with $\qv$ on $\Gamma$M($\Gamma$K) with a length given by Eqs.~\eqref{eq:QII}-\eqref{eq:QIII}. \markup{Examples of both the high-$T$ and low-$T$ structure factors near the phase transition for a generic parameter point in region $\RN{2}$($\RN{3}$) are shown in Fig.~\ref{fig:Keff_II}(Fig.~\ref{fig:Keff_III}).}

The structure factor is inherently inversion symmetric, and a \singleq state is thus characterized by \textit{two} peaks in the structure factor (both $\qv$ and -$\qv$). If one however considers one of these peaks alone, it will keep mirror symmetry about one \markup{of the $\Gamma$M lines} in regions $\RN{1}$ and $\RN{2}$, while it in region $\RN{3}$ keeps mirror symmetry about \markup{one of the $\Gamma$K lines.}

\subsection{$\RN{2}$--$\RN{3}$ border}
\begin{figure}[h!]
\begin{center}
\includegraphics[width=0.43\textwidth]{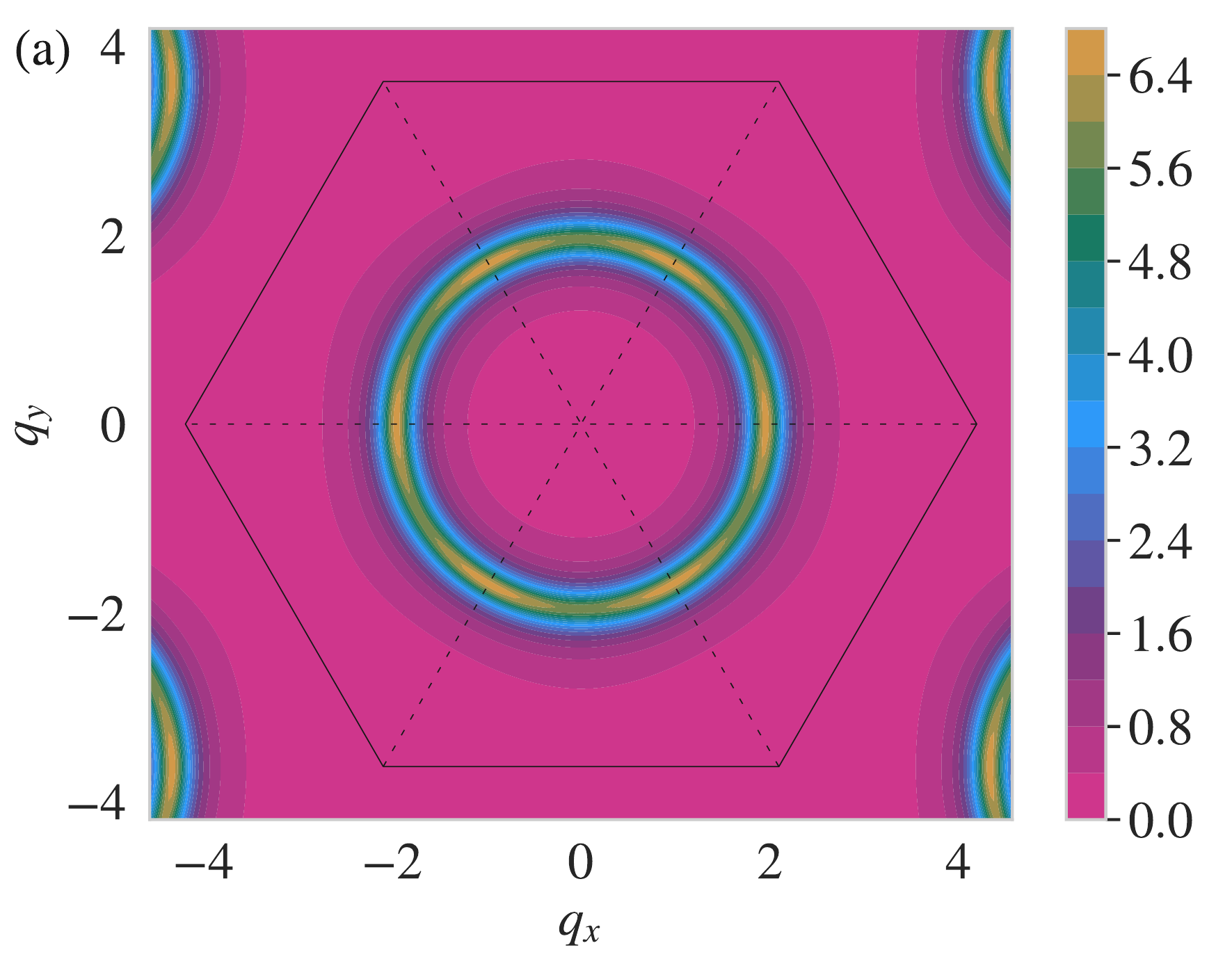}
\includegraphics[width=0.43\textwidth]{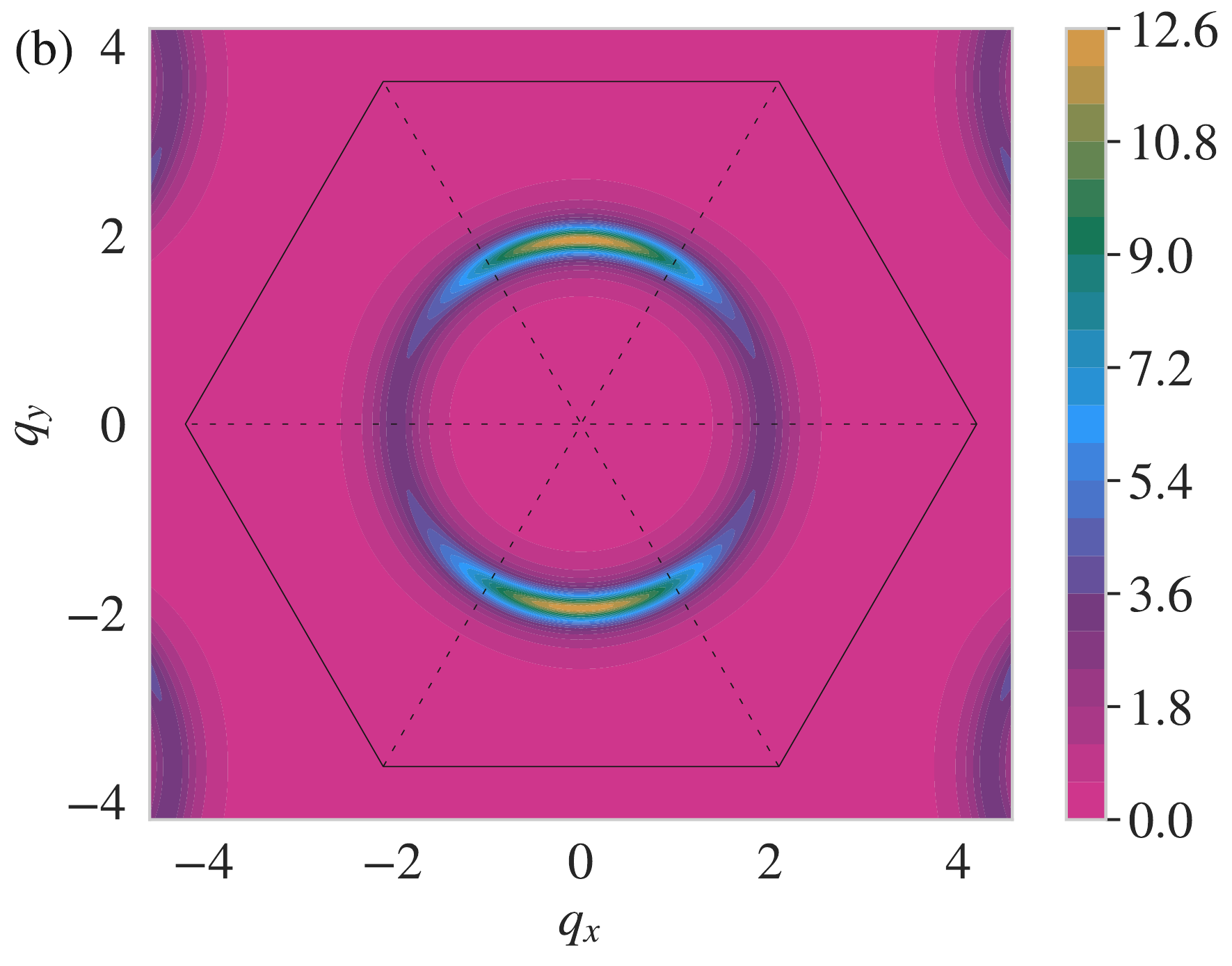}
\includegraphics[width=0.43\textwidth]{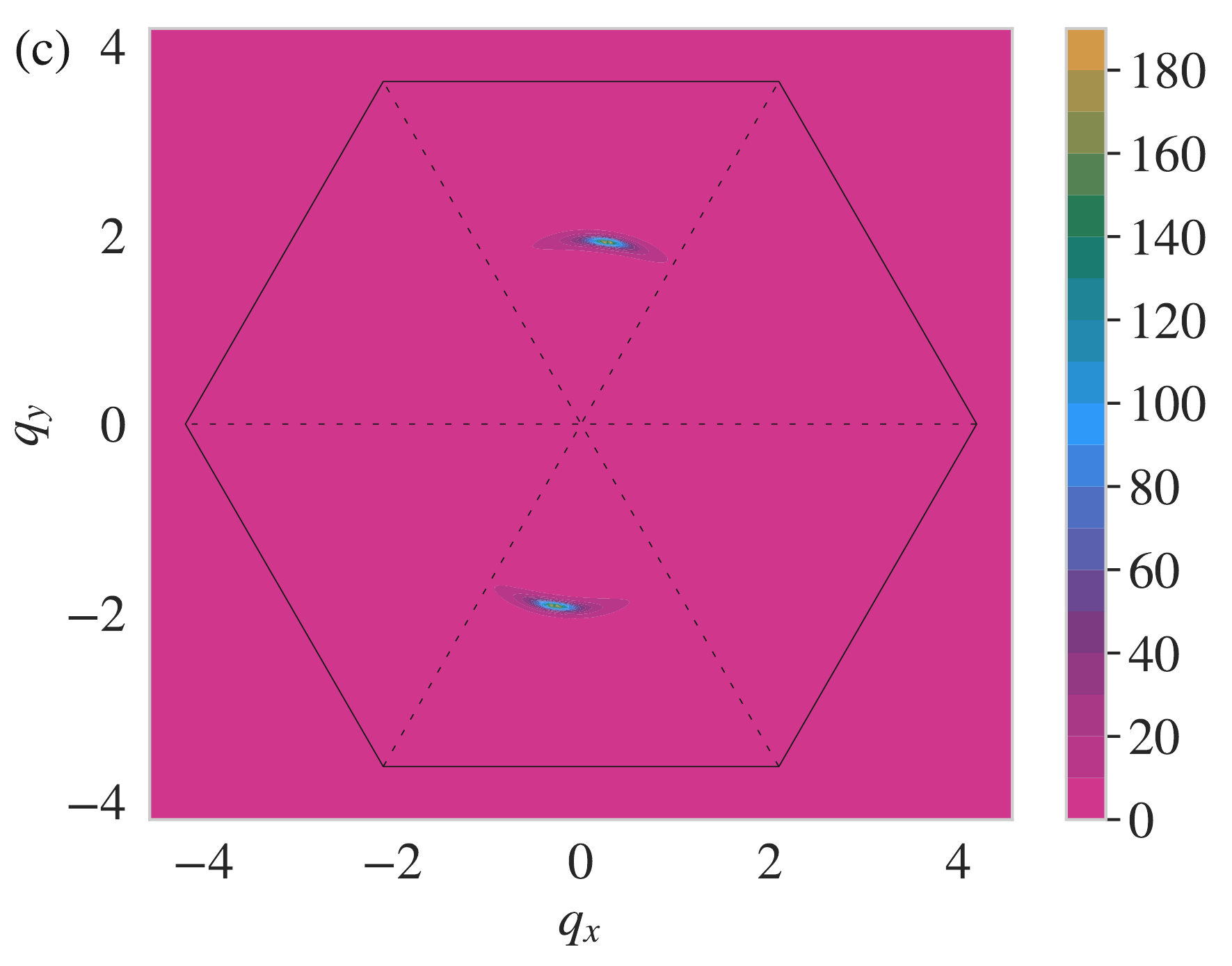}
\caption{Structure factors $\mathcal{S}(\qv)$ at the $\RN{2}$--$\RN{3}$ border, $(J_2,J_3) = (2,1)$, for different temperatures: (a) high-$T$ ring liquid phase at $T=0.943$, (b) intermediate arc \markup{regime} at $T=0.927$ and (c) low-$T$ non-symmetric phase at $T=0.788$. $L=200$.
\label{fig:structurefactors}}
\end{center}
\end{figure}

For parameter values near the $\RN{2}$--$\RN{3}$ border, on which the $\Qv$s form a continuous set, the phase structure is more complicated. In particular we find that exactly on the border, $J_2=2J_3$, there are {\em two} consecutive phase transitions as the temperature is lowered. Fig.~\ref{fig:structurefactors} shows the structure factors in the three distinct phases. At high-$T$ the system is in the fully symmetric ring liquid phase where the structure factor shows a ring, Fig.~\ref{fig:structurefactors}(a). Then below a first-order phase transition this ring is replaced by two partial rings/arcs, where only about one third of the full ring is present and centered on $\Gamma$M, Fig.~\ref{fig:structurefactors}(b). This \markup{arc structure factor} breaks rotational symmetry, but is mirror symmetric about $\Gamma$M. We describe this \markup{regime} in more detail in the following subsection. Then below this, there is a second phase transition into a single-$\qv$ non-symmetric phase where the structure factor has a narrow peak centered on a point $\qv^{\,*}$ which is neither along $\Gamma$M nor $\Gamma$K, see Fig.~\ref{fig:structurefactors}(c).  In fact, $\qv^{\,*}$ rotates continuously towards the value predicted by the maximum entropy of spin waves around single-$\qv$ spirals as the temperature is lowered, see appendix~\ref{app:Spin wave entropy}. This single-$\qv$ phase breaks all the lattice symmetries except inversion symmetry. The free energy is qualitatively similar to Fig.~\ref{fig:FreeE} and shows a first-order phase transition between the ring liquid and the arc \markup{regime}, but no apparent discontinuity in the derivative at the low-$T$ phase transition\markup{. The breaking of the remaining lattice mirror symmetries of phase $\RN{2}$ should however be accompanied by a phase transition, and }thus we conclude that the low-$T$ phase transition between the arc \markup{regime} and the non-symmetric phase is continuous. \markup{The transition temperature is in this case found by considering the symmetries of the structure factor.}

\begin{figure}[t]
\begin{center}
\includegraphics[width=0.45\textwidth]{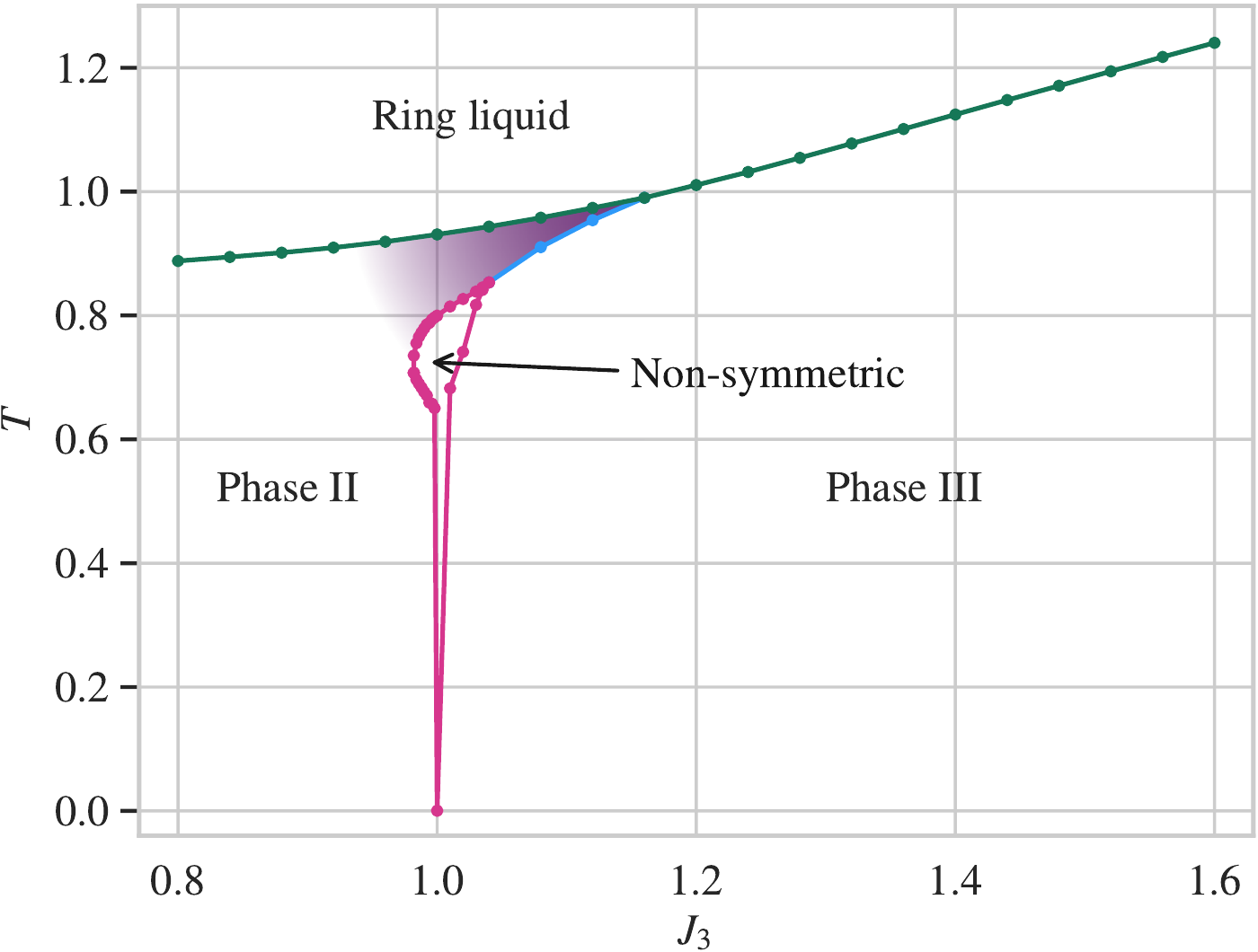}
\caption{Phase diagram for $J_2 = 2$. The purple region indicates the arc \markup{regime}. The green and blue curves indicate first-order phase transitions, while the pink curve indicates continuous phase transitions. \markup{The $\RN{2}$--$\RN{3}$ border is at $J_3=1$.}}
\label{fig:phase_J2-2}
\end{center}
\end{figure}

By investigating also $J_3$-values away from the $\RN{2}$--$\RN{3}$ border for $J_2=2$ we establish the phase diagram shown in Fig.~\ref{fig:phase_J2-2}. \markup{The phase diagram shows four} phases: At high-$T$, we find the ring liquid phase, where all lattice symmetries are present. Phase $\RN{2}$ and phase $\RN{3}$ break rotational symmetry while keeping some mirror symmetries. The non-symmetric phase is a single-$\qv$ state in which both the rotational symmetry and all the mirror symmetries are broken. Phase $\RN{2}$ and phase $\RN{3}$ are in general single-$\qv$ spiral states, where $\qv$ is determined by the respective minima of $J_{\qv}$.  \markup{The arc regime, discussed below, is shown in purple. This regime is continuously connected to phase $\RN{2}$, while a first-order phase transition separates it from phase $\RN{3}$.}

\subsection{\markup{Arc regime}}
\begin{figure}[t]
\begin{center}
\includegraphics[width=0.45\textwidth]{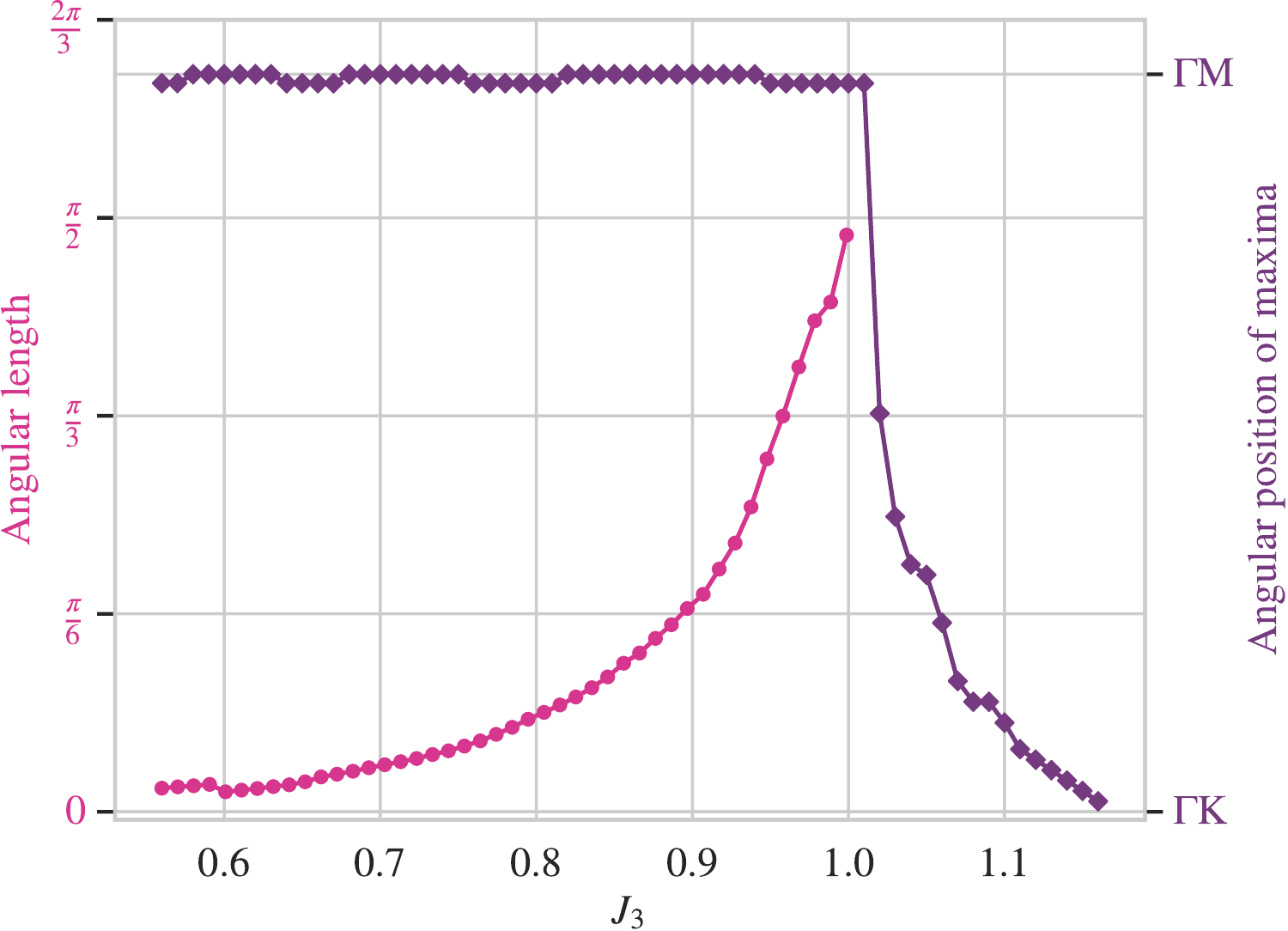}
\caption{Properties of the structure factor arc at the highest $T_c$ for $J_2= 2$. $L=500$. Pink circles: Angular length of the structure factor arc defined as the full width at half maximum. Purple diamonds: The angular position of the maximum/maxima of the structure factor. The arc is always centered on $\Gamma$M, thus there is a split maximum for $J_3 > 1$.}
\label{fig:width_theta_J2-2_L-500}
\end{center}
\end{figure}

The \markup{structure factor arc}, Fig.~\ref{fig:structurefactors}(b), has the same symmetries as the single-$\qv$ phase in region $\RN{2}$ where the peak is centered on $\Gamma$M. However, the \markup{structure factor arc} near the $\RN{2}$--$\RN{3}$ border cannot be characterized as a single-$\qv$ state as the arc length covers almost a quarter of the full circle. Fig.~\ref{fig:width_theta_J2-2_L-500} shows how the angular length of the arc and the position of its maximum change as the $\RN{2}$--$\RN{3}$ border is approached from the region $\RN{2}$ side. The arc length increases monotonically, while the maximum intensity is on $\Gamma$M.

\begin{figure}[t]
\begin{center}
\includegraphics[width=0.45\textwidth]{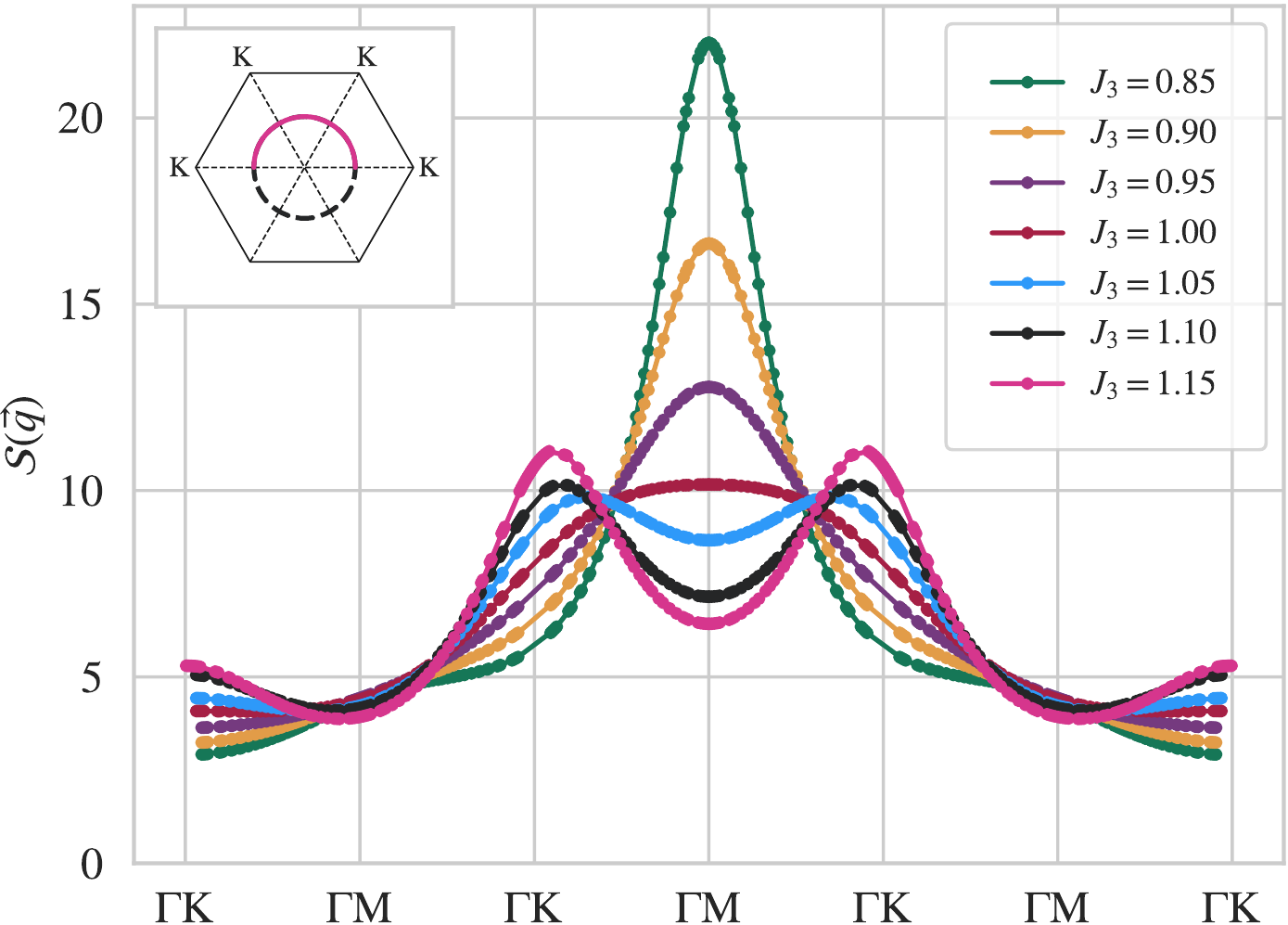}
\caption{The arc intensity at constant $\abs{\qv}$ for $T$ just below the highest $T_c$ and $J_2= 2$. $L=800$. The horizontal axis shows the angular variation of $\qv$, illustrated by the pink solid line in the inset.}
\label{fig:wall_splitting_vary_J3_J2-2_L-800}
\end{center}
\end{figure}

\begin{figure}[t]
\begin{center}
\includegraphics[width=0.45\textwidth]{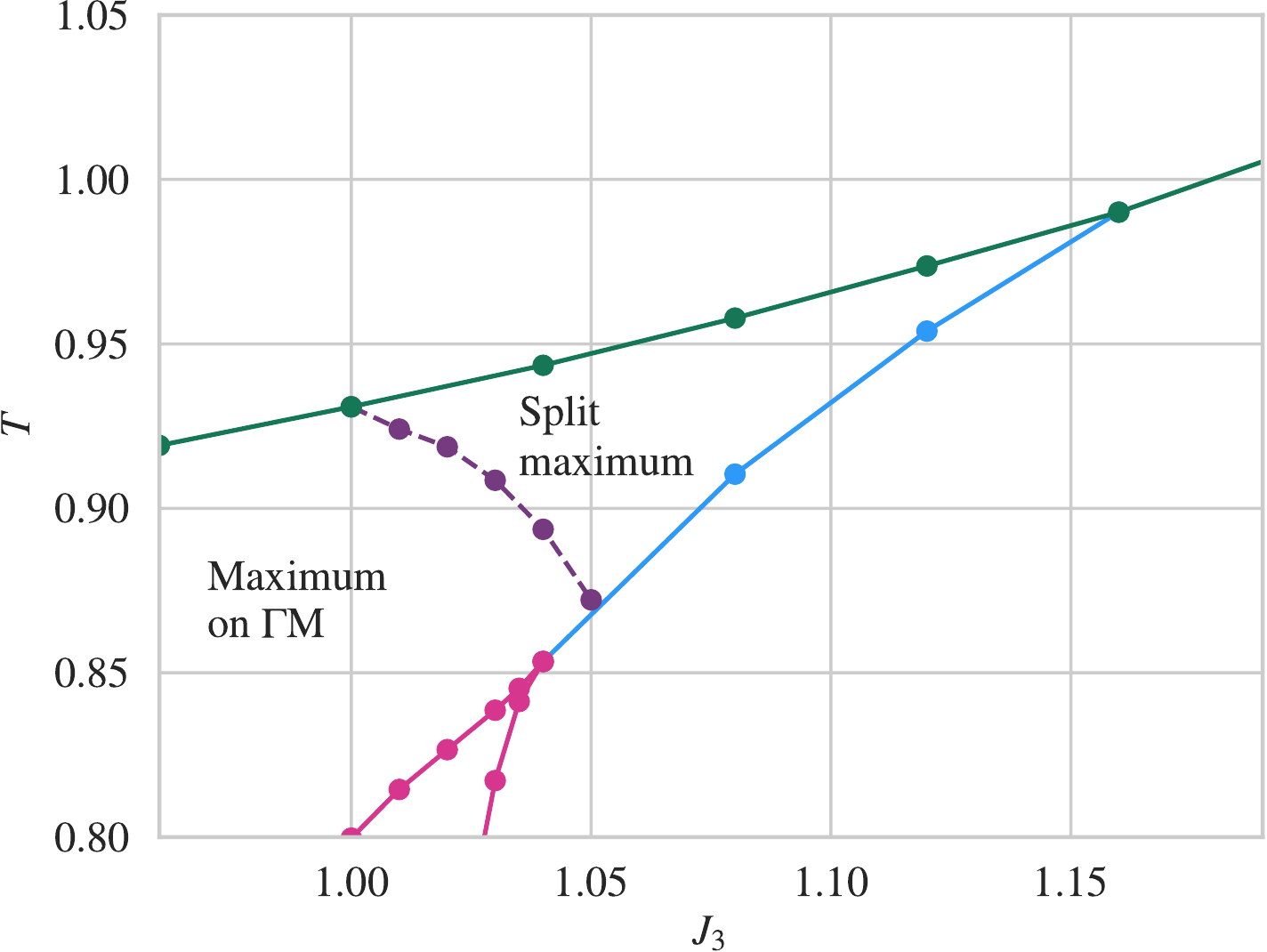}
\caption{Cut-out of Fig.~\ref{fig:phase_J2-2}. The purple dashed line shows where the maximum of the arc intensity goes from being on $\Gamma$M to splitting into two maxima located symmetrically about $\Gamma$M. $J_2= 2$.}
\label{fig:M_shape_J2-2}
\end{center}
\end{figure}

Intriguingly, we see from Fig.~\ref{fig:phase_J2-2} that the \markup{arc regime} (purple region) also extends into the region $\RN{3}$ side of the $\RN{2}$--$\RN{3}$ border where $J_{\qv}$ develops minima at $\Gamma$K. On this side, the arc intensity develops a split maximum with two peaks located symmetrically about $\Gamma$M. These peaks approach $\Gamma$K as $J_3$ is increased, as seen for $J_3 > 1$ in Fig.~\ref{fig:width_theta_J2-2_L-500}. Examples of the arc intensity just below the highest $T_c$ for different $J_3$ are shown in Fig.~\ref{fig:wall_splitting_vary_J3_J2-2_L-800}. These intensity shapes depend also on the temperature: When lowering the temperature from $T_c$, the split peaks move towards $\Gamma$M. Fig.~\ref{fig:M_shape_J2-2} shows where the arc intensity has its maximum on $\Gamma$M and where the maximum is split.

\begin{figure}[t]
\begin{center}
\includegraphics[width=0.45\textwidth]{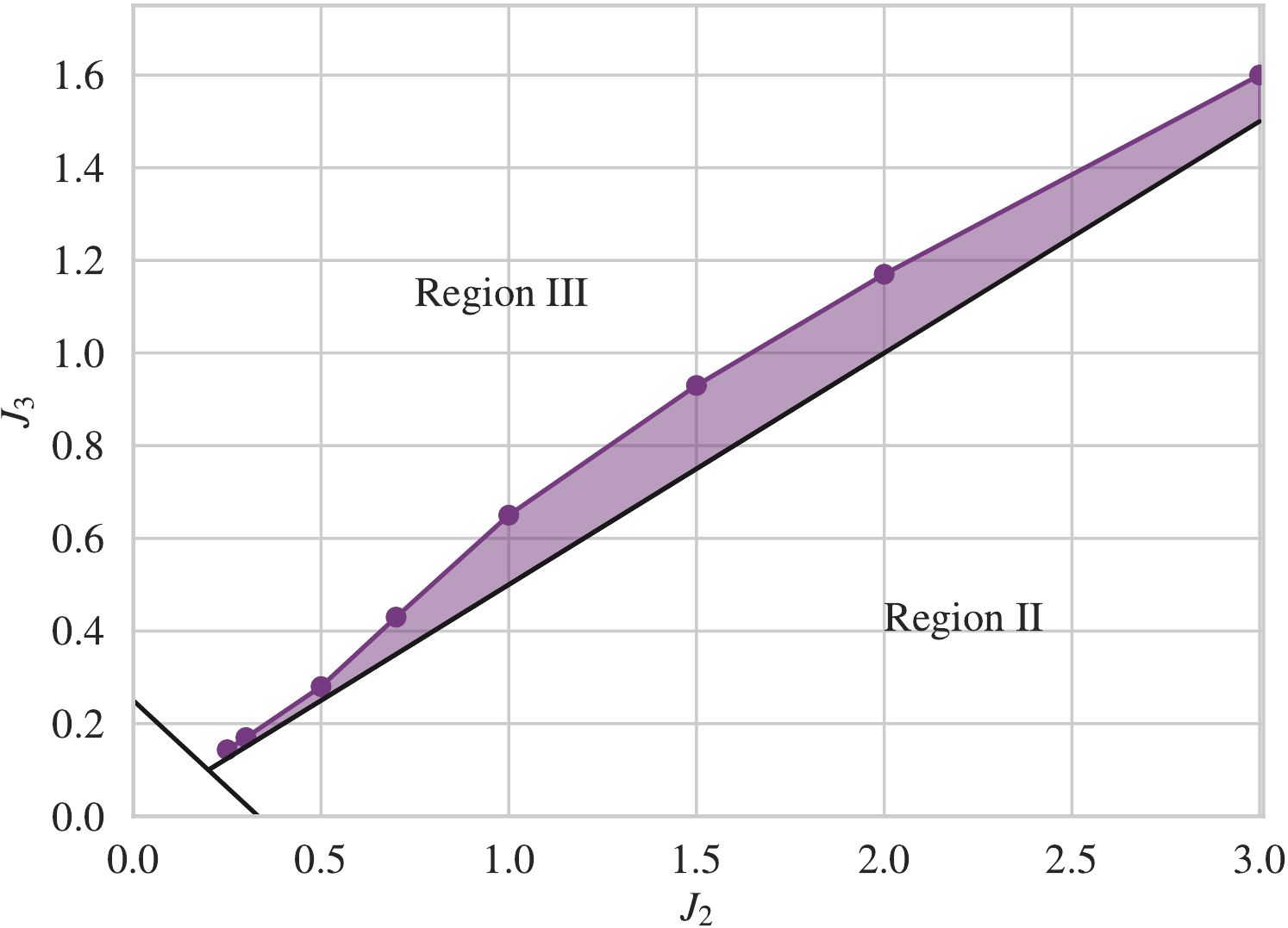}
\caption{An illustration of where in parameter space the \markup{arc regime} exists. We have only studied $J_2 \leq 3$. The \markup{arc regime} also exists on the region $\RN{2}$ side of the $\RN{2}$--$\RN{3}$ border. \markup{However, as the arc regime is continuously connected to phase $\RN{2}$, it cannot be distinguished from phase $\RN{2}$ in a well-defined way, as shown for $J_2 = 2$ in Fig.~\ref{fig:phase_J2-2}.}}
\label{fig:J2J3_doublePT}
\end{center}
\end{figure}

The \markup{arc regime} exists also for other values of $J_2$ near the $\RN{2}$--$\RN{3}$ border, see Fig.~\ref{fig:J2J3_doublePT}.

\section{Discussion}\label{sec:discussion}

The behavior of the {\Jonetwothree} Heisenberg model on the triangular lattice with $J_1 < 0$ is well-understood for the whole parameter space at low temperatures: when the $\Qv$s form a discrete set in regions $\RN{1}$, $\RN{2}$ and $\RN{3}$, the system breaks lattice rotational symmetry by forming one of the \singleq spiral states with minimal energy. Wherever the $\Qv$s form a continuous set, i.e. at the $\RN{2}$--$\RN{3}$ border, the degeneracy is lifted by the spin wave entropy, and the system breaks lattice rotational symmetry by forming one of the \singleq spiral states with maximal entropy.

In the discrete case, we find that the transition into the low-$T$ ordered phase from the high-$T$ symmetric state is generally a direct first-order phase transition. This is in agreement with Monte Carlo simulations on the triangular {\Jonethree} model.\cite{Tamura2008,Tamura2011} The critical temperatures obtained in this paper are however likely to be overestimated \markup{as was seen in Ref.~\onlinecite{Syljuasen2019} for layered square lattices. We believe this is caused by the neglect of fluctuations associated with vertex-corrections in the NBT.} \markup{We have made sure that all of our results are carried out at a sufficiently large system size, so that increasing it only gives minor corrections.}

Close to the $\RN{2}$--$\RN{3}$ border the phase transition is not direct. Instead, as the temperature is lowered from the high-$T$ phase, there is a first-order phase transition to an intermediate \markup{regime: the arc regime}. Then at lower $T$ there is a second transition. If the system is at or very close to the $\RN{2}$--$\RN{3}$ border, this second transition is a continuous phase transition into the non-symmetric  \singleq phase. In this phase, the pitch vector of the spiral changes continuously as $T$ is further lowered and reaches eventually the value maximizing the spin wave entropy. Further into the region $\RN{3}$ side of the $\RN{2}$--$\RN{3}$ border, the second phase transition becomes first-order into the \singleq phase $\RN{3}$.
Such two-step pattern of symmetry-breaking vaguely resembles the well-known hexatic melting scenario where the system with broken orientational and translational order is melted via an intermediate hexatic phase which breaks translational, but not orientational, order.\cite{HalperinNelson78,NelsonHalperin79}

The \markup{structure factor arc has the same symmetries as the \singleq states in phase $\RN{2}$.} Nevertheless the \markup{structure factor arc} cannot be characterized as a \singleq state. In fact, the static structure factor of the \markup{arc} resembles rather the high-$T$ ring liquid, but with portions of the ring removed. If one interprets the ring liquid as a \markup{spiral liquid consisting of a} collection of short spirals with pitch vectors free to point in any direction, but constrained to have magnitudes lying on the manifold $\Qv$, it is natural to conjecture that the \markup{arc} is similar, but
with the orientation of the spiral pitch vectors restricted to a distribution about one $\Gamma$M. This would also explain the split maximum of the arc intensity towards $\Gamma$K, as domains with \singleq spirals along $\Gamma$K become energetically favorable on the region $\RN{3}$ side of the $\RN{2}$--$\RN{3}$ border. However, the coexistence of many spiral\markup{s} is far from trivial, and leads naturally to the consideration of energy and entropy of domain walls between \singleq spiral domains. A very impressive characterization and observation of these has recently been done for helical magnets where the pitch vector is perpendicular to the spiral plane.\cite{Li2012, Schoenherr2018, Nattermann2018} In particular bisector domain walls, where the domain wall bisects the two pitch vector orientations on either side, are favorable energetically. The analysis of how such domain walls lead to phase transitions must also include their entropy induced by kinks and spin waves.  Such an analysis for the {\em Ising} model with extended range interactions on the triangular lattice showed that double domain walls\cite{Korshunov2005} lead to an intermediate nematic phase.\cite{Smerald2016}
We note that stable {\em point} defects can also exist in triangular lattice antiferromagnets,\cite{KawamuraMiyashita1984} but their role in breaking lattice symmetries is unclear.
In order to understand the \markup{arc regime}, lattice details must also be accounted for to explain why the arc is centered \markup{on $\Gamma$M}, and not \markup{on $\Gamma$K.}

The \markup{structure factor arc} resembles strikingly the half-moon patterns seen in simulations~\cite{Robert2008} and experiments~\cite{Guitteny2013} on kagome and pyrochlore lattices. These half-moons occur both in the dynamic\cite{Yan2018} and static structure factors,\cite{Mizoguchi2018} however they do not break lattice rotational symmetry. Furthermore, the static half-moons are a consequence of having several atoms in the unit cell, as the half-moon is the complement of the flat band combined with another dispersive band with a continuous minimum.\cite{Mizoguchi2018} Thus, except for their appearance, it is not clear if or how the \markup{structure factor arc} is related to the half-moons.

It is pertinent to contrast the result obtained here to that obtained for the {\Jonetwothree} Heisenberg model on the square lattice. Among other results, Ref.~\onlinecite{Seabra2016} found an intermediate vortex crystal phase between the \singleq and the ring liquid at a single parameter point where the $\Qv$s form a continuous set. The vortex crystal state has a structure factor peaked on four particular momentum vectors. Such a state is favorable when all these four momentum vectors lie at or very near the minimal $J_{\qv}$ contour. We have attempted construction of similar combinations of 3-$\qv$ and 4-$\qv$ states for the triangular lattice {\Jonetwothree} model, but have not found a suitable candidate that keeps the spins normalized, and where all the $\qv$\;'s minimize $J_{\qv}$ simultaneously. In any case, if there is such a candidate, the resulting vortex crystal would probably only exist in a narrow range about one particular parameter point, and not for such an extended region in parameter space as we have found the \markup{arc regime}.

\markup{To strengthen the validity of our findings, it would be very valuable if our results could be confirmed by independent Monte Carlo simulations.
We also }hope that future research will properly explain the origin of the \markup{arc regime}. Experimentally, the results obtained in this paper should be relevant for any magnetic material that can be described by the classical {\Jonetwothree} Heisenberg model on the triangular lattice with ferromagnetic $J_1$. In such materials, the first-order magnetic lattice symmetry breaking phase transitions, that we have found to occur over large portions of the phase diagram, may also be accompanied by concomitant structural instabilities triggered through magnetoelastic couplings.\cite{Fang2008} An experimental observation of the \markup{arc regime} will probably have to await a genuinely tunable magnet where the coupling parameters can be adjusted so that the minimal $\Qv$s form a continuous set. We note that a magnetic system with tunable anisotropy has already been realized with cold atoms.\cite{Jepsen2020}

\markup{The NBT method used here can also be employed to investigate other spiral liquids, such as the extended Heisenberg model on the diamond lattice, relevant for the material \ce{MnSc2S4}.\cite{Bergman2007}}

\acknowledgments{
OFS acknowledges stimulating discussions with Jens Paaske.
The computations were performed on resources provided by UNINETT Sigma2 - the National Infrastructure for High Performance Computing and Data Storage in Norway.}

\appendix

\section{\Srest \label{app:Sremainder}}
In Eq.~\eqref{SofDelta}, $S[\Delta]$ is expressed in terms of renormalized propagators and a remainder
\begin{align}
\Srest &\equiv -\f{N_s}{2} \sum_{n=1}^\infty \f{n-1}{n} \tr \left( \Kinvmat \Sigmamat \right)^n + \f{1}{2} \sum_{n=1}^{\infty} \f{1}{n} \tr \left( \Dmat_0 \Pimat \right)^n
\nonumber \\
&\quad
- \ln \langle e^{-S_r} \rangle,
\label{Srest}
\end{align}
where it is understood in Eq.~\eqref{SofDelta} that the $\qv=0$ contribution must be omitted when evaluating $\tr \ln{\Dinvmat}$ and $\tr \left(\Dmat_0 \Pimat \right)^n$.
In arriving at this expression we have added and subtracted a term
\be
\f{N_s}{2} \tr \left( \Keffinvmat \Sigmamat \right) = \f{N_s}{2} \sum_{n=1}^\infty \tr \left( \Kinvmat \Sigmamat \right)^n
\ee
so as to cancel the term $\tr (\Kinvmat \Sigmamat)$ in \Srest. This term causes $\Srest$ to be $\order{N_s^0}$ as there are no single--ring diagrams with one wavy line in $\ln{\langle e^{-S_r} \rangle}$.
The term $\ln{\langle e^{-S_r} \rangle}$ can be evaluated using the cumulant expansion, and consists of all connected diagrams of rings with three or more wavy hooks. Each wavy line carries a factor $D$ which is $\order{1/N_s}$ and each ring with $n$ wavy hooks a factor $N_s(-i)^n/2n$ and $n$ factors of $\Kinv$. Momentum is conserved at every vertex.

Many diagrams cancel each other in $\Srest$. In particular the types shown in Fig.~\ref{includeddiagrams}. To see this, take first the connected diagram with $m \geq 1$ identical rings each with $2k$ hooks contracted sequentially in the fashion shown in Fig.~\ref{includeddiagrams} left for the case $k=2$. The $m$'th cumulant of $-\ln{\langle e^{-S_r} \rangle}$ gives this diagram with a combinatorial factor $-(-1)^{km} (1/2)^{sm}/2m$ where $s$ is a symmetry factor which is 1 if the ring with $2k$ hooks is symmetric when flipped about its external wavy lines and zero otherwise.
\mycomment{
\\
For a single ring with $2k$ hooks there are $2k$ possibilities for selecting an entrance leg. Then there are 2 ways of contracting neighboring hooks to  a particular pattern, however if the diagram is symmetric there is only one way, thus a factor $(1/2)^s$, where $s=1$ is the diagram is symmetric. This procedure overcounts the diagrams by 2 because the exit leg could also have been chosen as entrance leg. Thus there are in total $2k (1/2)^s$ ways of contracting the labelled vertices to the same diagram. In addition there is a multiplicative factor $N_s (-i)^{2k}/2(2k)$ associated with this diagram.

Then we join all the labelled legs on the $m$ rings together, this can be done in $(m-1)! 2^m/2$ different ways, and when multiplying by the overall $1/m!$ factor from the cumulant we get:
\be
-\f{1}{m!} \f{(m-1)! 2^m}{2} \left[ \f{N_s (-i)^{2k}}{2(2k)} 2k \left( \f{1}{2} \right)^s \right]^m  = -\f{1}{2m} \left[ N_s \left( \f{1}{2} \right)^{s} (-1)^k \right]^m  \nonumber
\ee
}
The term $\f{1}{2m} \tr \left( \Dzeromat \Pimat \right)^m$ gives also this diagram when $\Pimat$ is expanded to the $k-1$'th order in the self-energy. In fact, it gives the same contribution, but with opposite sign.
\mycomment{
\\
The polarization can be written symbolically as
\begin{align}
\Pi &= -\f{N_s}{2} \left\{ \left[ \left( 1-\Kinv \Sigma \right)^{-1} \Kinv
,\left( 1-\Kinv \Sigma \right)^{-1} \Kinv \right] \right. \nonumber \\
& \qquad \left. - \left[ \Kinv, \Kinv \right] \right\} \nonumber
\end{align}
where the brackets means $[A,B] \equiv \sum_{\pv} A_{\pv+\qv} B_{\pv}$.
Expanding this as power series in $\Sigma$ we get in particular a term with $a+b=k-1$ powers of $\Sigma$ which is
\be
\Pi = - N_s \left(\f{1}{2} \right)^s \left[ \left( \Kinv \Sigma \right)^a \Kinv
,\left( \Kinv \Sigma \right)^b \Kinv \right] \nonumber
\ee
where the symmetry factor arises because there is only one term if $a=b$, not two as for the cases where $a \neq b$.
Now we can use the lowest order expression for $\Sigma$, $\Sigma \approx - [\Kinv, D_0]$ which gives the desired ring diagram with a factor
\be
- N_s \left(\f{1}{2} \right)^s \left( -1 \right)^{k-1} =
N_s \left(\f{1}{2} \right)^s \left( -1 \right)^{k} \nonumber
\ee
Since all the rings are identical, all the polarizations must contribute exactly this term, and the overall combinatorial factor becomes
\be
\f{1}{2m} \left[ N_s \left(\f{1}{2} \right)^s \left( -1 \right)^{k} \right]^m \nonumber
\ee
}
In the cases when the external hooks on each ring are nearest neighbors, like the diagram Fig.~\ref{includeddiagrams} left, there are additional contributions.
The first comes from the term $-\f{N_s(k-1)}{2k} \tr \left( \Kinvmat \Sigmamat \right)^{k}$ where one of the $\Sigmamat$ is written in terms of the full propagator $\Dmat$ which in turn is expanded to the $m-1$'th power in the polarization, while the rest are replaced with its lowest order contribution. This gives the combinatorial factor $-(-1)^{k}(k-1)/2$.
\mycomment{
\\
There are $k$ choices for which self-energy to expand, thus there is an extra factor $k$. Also there is a minus sign for each self-energy.
\begin{align}
-\f{N_s (k-1)}{2k} k (-1)^k  & \tr \left( \Kinv \left[\Kinv, D_0\right] \right)^{k-1} \Kinv \left[ \Kinv, D \right]
\end{align}
Then we take the $m-1$'th term in the expansion of $D \sim \left( D_0 \Pi \right)^{m-1} D_0$ which gives the desired diagram
with the combinatorial factor
\be
-\f{N_s(k-1)}{2} (-1)^k
\ee
}
The second contribution, which cancels the first, comes from the term $\f{1}{2} \tr \left( \Dzeromat \Pimat \right)$ when expanding the polarization in terms of the self-energy to the $k-1$'th power, and then replacing one of the self-energies with the full propagator $D$ and the rest with $D_0$.
\mycomment{
\\
\be
\f{1}{2} \tr D_0 \Pi = -\f{N_s}{2} \tr D_0 \left[ \Kinv, \left( \Kinv \Sigma \right)^{k-1} \Kinv \right]
\ee
Since we are taking the trace we can use the identity
\be
\tr A [B,C] = \tr C [B,A]
\ee
which holds because
\be
\sum_{\qv} A_{\qv} \sum_{\pv} B_{\qv+\pv} C_{\pv} =
\sum_{\qv,\pv} A_{\qv} B_{\qv+\pv} C_{\pv} =
\sum_{\pv} C_{\pv} \sum_{\qv} B_{\qv+\pv} A_{\qv}
\ee
and gives
\be
\f{1}{2} \tr \left( D_0 \Pi \right)
= -\f{N_s}{2}  \tr \left( \Kinv \Sigma \right)^{k-1} \Kinv \left[ \Kinv,D_0 \right]
\ee
Then we write one of the $\Sigma$'s using $D$ and replace the rest with the lowest order approximation $D_0$. This gives
\begin{align}
\f{1}{2} \tr \left( D_0 \Pi \right) &= -\f{N_s}{2} (-1)^{k-1} (k-1) \tr \left( \Kinv \left[ \Kinv,D_0 \right] \right)^{k-2} \nonumber \\
& \qquad  \times \Kinv \left[ \Kinv,D \right] \Kinv \left[ \Kinv,D_0 \right]
\end{align}
using the trace cyclicity this can be rewritten as
\begin{align}
\f{1}{2} \tr \left( D_0 \Pi \right) &= \f{N_s(k-1)}{2} (-1)^{k} \tr \left( \Kinv \left[ \Kinv,D_0 \right] \right)^{k-1} \nonumber \\
& \qquad  \times \Kinv \left[ \Kinv,D \right]
\end{align}
Now expanding $D$ to $m-1$'th power we see that we get the same as above, but with opposite sign. In fact, already before expanding we see that these two terms cancel.
}
Therefore all these diagrams vanish in $\Srest$.
Similarly the single ring diagram with m sequential wavy lines shown in Fig.~\ref{includeddiagrams} right will also vanish. Adding together the combinatorial factors: $-1/2m + 1/2 - (m-1)/2m $ that comes from the terms $-\ln{\langle e^{-S_r} \rangle}$, $\f{1}{2} \tr \left( \Dzeromat \Pimat \right)$ and $-\f{N_s(m-1)}{2m} \tr\left( \Kinvmat \Sigmamat \right)^m$ respectively, we get zero.

\section{Saddle point \label{app:Saddlepoint}}
The saddle point method including Gaussian corrections gives
\be
\int (-i) d\Delta e^{-S[\Delta]} \propto e^{-S[\Delta_0] - \f{1}{2} \ln{\left(-\f{\partial^2 S[\Delta]}{\partial \Delta^2} \right)}|_{\Delta=\Delta_0}},
\ee
where the saddle point value of $\Delta$ is determined by setting $\f{\partial S}{\partial \Delta}=0$. We have here restored the factor $-i$ which comes from changing the variable $\lambda_{\qv=0}= -i\Delta$.
Differentiating $S[\Delta]$ gives
\begin{align}
\f{\partial S[\Delta]}{\partial \Delta} &= -\beta V + \f{N_s}{2} \tr \left( \Keffinvmat \left( 1 - \f{\partial \Sigmamat}{\partial \Delta} \right) \right) + \f{1}{2} \tr \Dmat \f{\partial \Dinvmat}{\partial \Delta} \nonumber \\
& \quad +\f{N_s}{2} \tr \left( \f{\partial \Keffinvmat}{\partial \Delta} \Sigmamat + \Keffinvmat \f{\partial \Sigmamat}{\partial \Delta} \right).
\end{align}
This can be simplified by using Eq.~\eqref{constraintprop_eq} to deduce
\begin{align}
\f{\partial D^{-1}_{\qv}}{\partial \Delta} &= N_s \sum_{\pv} \Keffinv_{\pv+\qv} \f{\partial \Keffinv_{\qv}}{\partial \Delta},
\end{align}
and Eq.~\eqref{selfenergy_eq} to rewrite the self-energy. It follows that
\be
\f{\partial S[\Delta]}{\partial \Delta} = -\beta V + \f{N_s}{2} \tr \Keffinvmat,
\ee
which implies the saddle point condition Eq.~\eqref{saddlepoint}.

The second derivative is
\begin{align}
\f{\partial^2 S[\Delta]}{\partial \Delta^2} &= -\f{N_s}{2} \tr \left( \Keffinvmat \Keffinvmat \left( 1 - \f{\partial \Sigmamat}{\partial \Delta} \right) \right) \approx -D^{-1}_{\qv=0},
\end{align}
where we have neglected $\f{\partial \Sigmamat}{\partial \Delta}$ as it is $\order{1/N_s}$. The Gaussian correction to the saddle point gives thus the missing $\qv=0$ part of the $\ln{D_{\qv}}$--sum in the free energy.

\section{Spin wave entropy \label{app:Spin wave entropy}}

\begin{figure}[t]
\begin{center}
\includegraphics[width=0.45\textwidth]{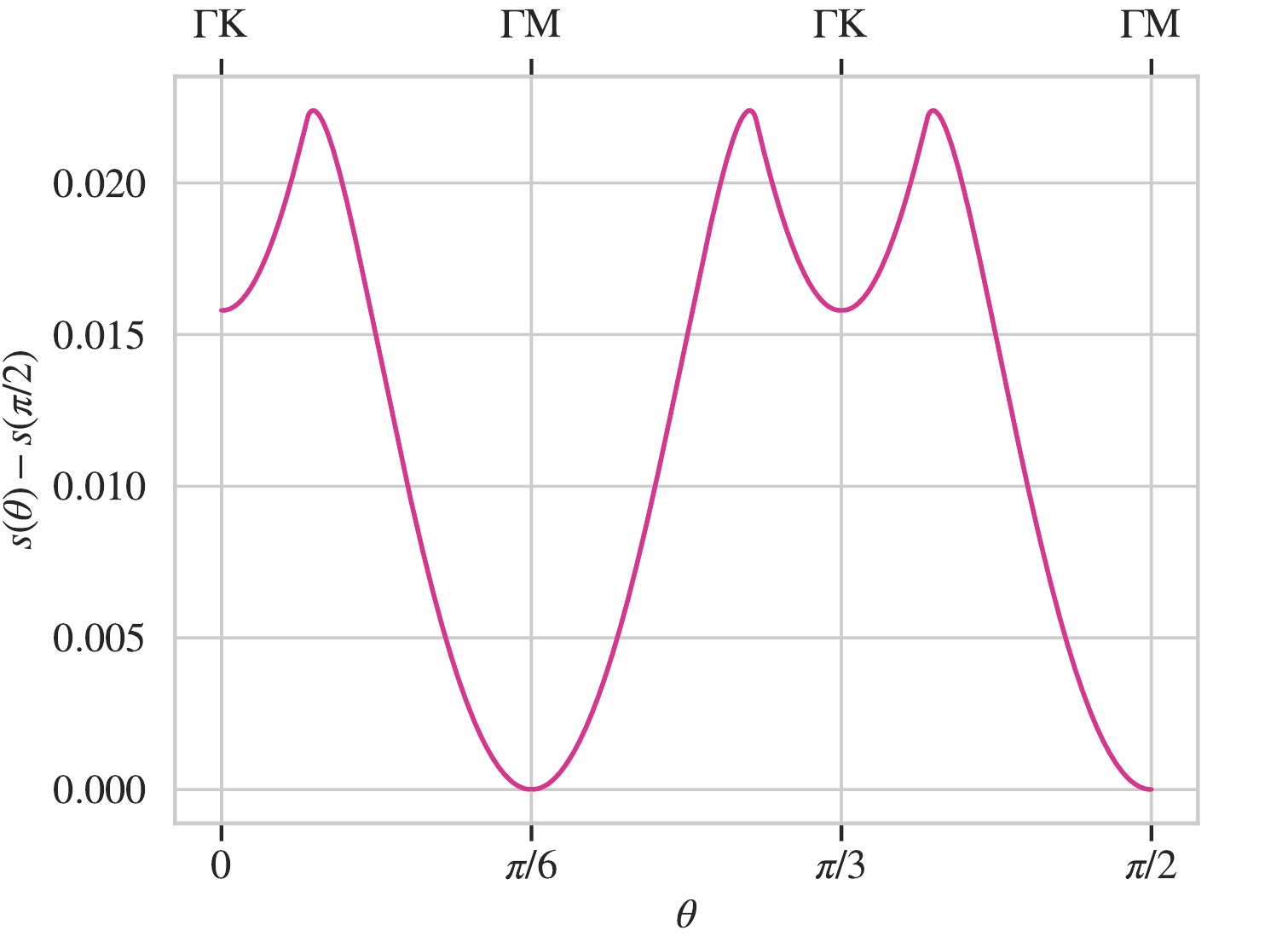}
\caption{Entropy of spin wave fluctuations around a\markup{n ordered} single-$\qv$ spiral state for $(J_2, J_3) = (2,1)$. The spiral pitch vector $\qv$ lies along ring-minimum with an angle $\theta$ relative to the $q_x$-axis.}
\label{fig:spiralentropy}
\end{center}
\end{figure}

\begin{figure}[t]
\begin{center}
\includegraphics[width=0.45\textwidth]{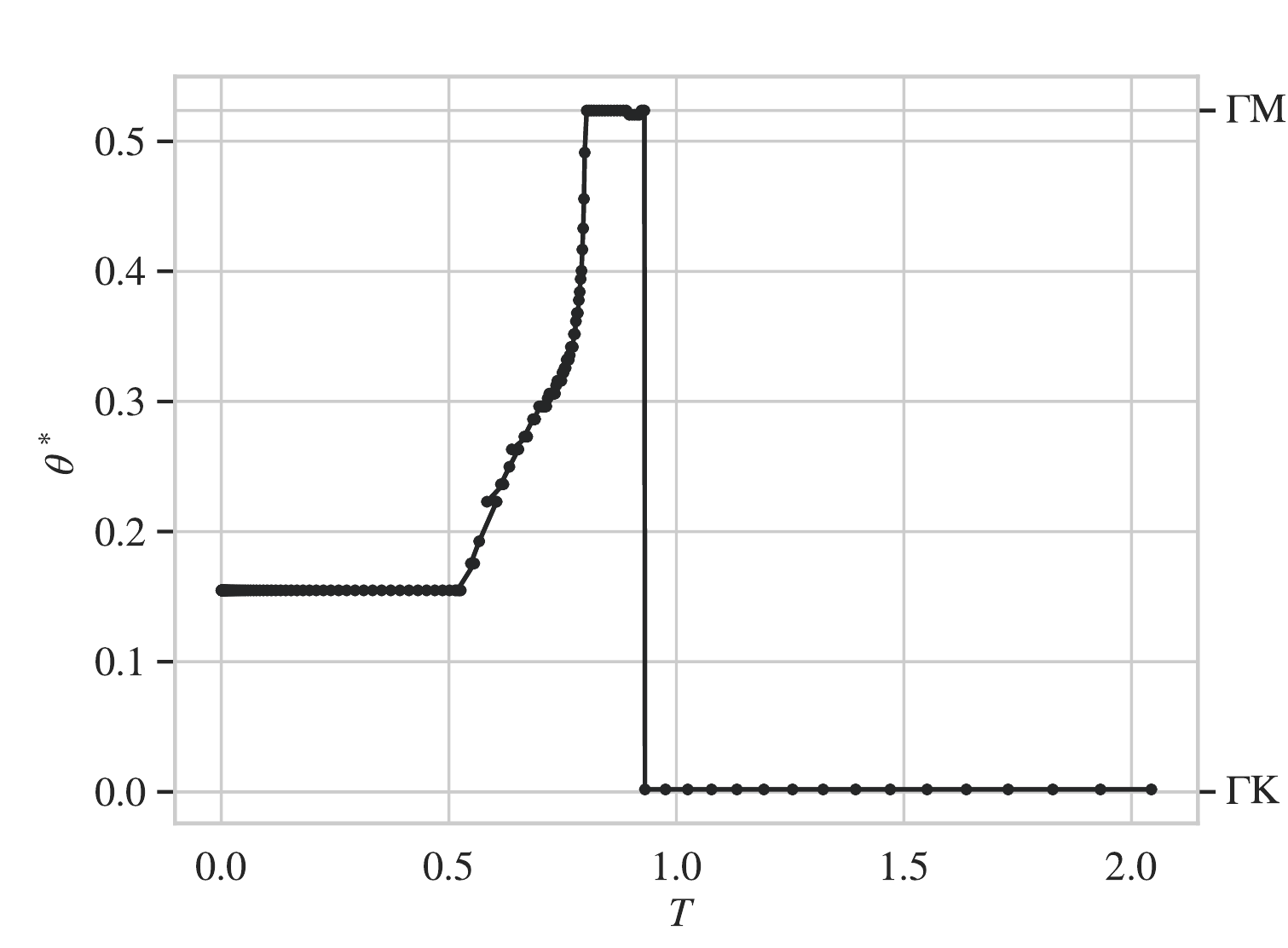}
\caption{Angular position of the maximum of $\mathcal{S}(\qv)$ as function of $T$ on the $\RN{2}$--$\RN{3}$ border, $(J_2, J_3) = (2,1)$. $L = 1000$.}
\label{fig:skeivtheta}
\end{center}
\end{figure}

The spin wave entropy per spin is given by
\begin{equation}\label{eq:spinwaveentropy}
s = -\frac{1}{V}\sum_{\vec{k}} \ln \beta \omega_{ \vec{k} },
\end{equation}
where $\omega_{ \vec{k} }$ is the spin wave dispersion. As shown in Ref.~\onlinecite{Seabra2016}, the spin wave dispersion around a\markup{n ordered} planar \singleq spiral state characterized by a pitch vector $\Qv$ is
\begin{equation}\label{eq:spinwavedispersion}
\omega_{ \vec{k}  } = \sqrt{\frac{1}{2}\left[ J_{ \Qv + \vec{k} } + J_{ \Qv - \vec{k} } - 2 J_{ \Qv } \right]\left[J_{ \vec{k} } - J_{ \Qv }\right]}.
\end{equation}
\markup{This result is strictly only valid where there is true long-range magnetic order at $T=0$. We assume here that we can nevertheless use it to find the $\Qv$s with maximum entropy also at low-T where the spin correlation length is large.}
The spin wave entropy for the minimal $\Qv$s when $(J_2, J_3) = (2,1)$ is shown in Fig.~\ref{fig:spiralentropy}. We have taken the entropy to be zero at $\Gamma$M. Taking $\theta$ to be the angle between $\Qv$ and the $q_x$-axis, we find that the spin wave entropy has its maximum for $\theta = 0.154$ and symmetry-related values.

Fig.~\ref{fig:skeivtheta} shows how the position $\qv^{\,*}$ of the maximum of $\mathcal{S}(\qv)$ varies with temperature for $(J_2, J_3) = (2,1)$. In the non-symmetric phase ($T < 0.80$), the angular position $\theta^*$ of $\qv^{\,*}$ changes continuously from $\theta^* = \pi/6$ ($\Gamma$M) to $\theta^* = 0.155$, which is very close to the value predicted by the spin wave entropy.


\bibliography{tri.bbl}

\begin{thebibliography}{34}%
\makeatletter
\providecommand \@ifxundefined [1]{%
 \@ifx{#1\undefined}
}%
\providecommand \@ifnum [1]{%
 \ifnum #1\expandafter \@firstoftwo
 \else \expandafter \@secondoftwo
 \fi
}%
\providecommand \@ifx [1]{%
 \ifx #1\expandafter \@firstoftwo
 \else \expandafter \@secondoftwo
 \fi
}%
\providecommand \natexlab [1]{#1}%
\providecommand \enquote  [1]{``#1''}%
\providecommand \bibnamefont  [1]{#1}%
\providecommand \bibfnamefont [1]{#1}%
\providecommand \citenamefont [1]{#1}%
\providecommand \href@noop [0]{\@secondoftwo}%
\providecommand \href [0]{\begingroup \@sanitize@url \@href}%
\providecommand \@href[1]{\@@startlink{#1}\@@href}%
\providecommand \@@href[1]{\endgroup#1\@@endlink}%
\providecommand \@sanitize@url [0]{\catcode `\\12\catcode `\$12\catcode
  `\&12\catcode `\#12\catcode `\^12\catcode `\_12\catcode `\%12\relax}%
\providecommand \@@startlink[1]{}%
\providecommand \@@endlink[0]{}%
\providecommand \url  [0]{\begingroup\@sanitize@url \@url }%
\providecommand \@url [1]{\endgroup\@href {#1}{\urlprefix }}%
\providecommand \urlprefix  [0]{URL }%
\providecommand \Eprint [0]{\href }%
\providecommand \doibase [0]{http://dx.doi.org/}%
\providecommand \selectlanguage [0]{\@gobble}%
\providecommand \bibinfo  [0]{\@secondoftwo}%
\providecommand \bibfield  [0]{\@secondoftwo}%
\providecommand \translation [1]{[#1]}%
\providecommand \BibitemOpen [0]{}%
\providecommand \bibitemStop [0]{}%
\providecommand \bibitemNoStop [0]{.\EOS\space}%
\providecommand \EOS [0]{\spacefactor3000\relax}%
\providecommand \BibitemShut  [1]{\csname bibitem#1\endcsname}%
\let\auto@bib@innerbib\@empty
\bibitem [{\citenamefont {Mermin}\ and\ \citenamefont
  {Wagner}(1966)}]{MerminWagner1966}%
  \BibitemOpen
  \bibfield  {author} {\bibinfo {author} {\bibfnamefont {N.~D.}\ \bibnamefont
  {Mermin}}\ and\ \bibinfo {author} {\bibfnamefont {H.}~\bibnamefont
  {Wagner}},\ }\href {\doibase 10.1103/PhysRevLett.17.1133} {\bibfield
  {journal} {\bibinfo  {journal} {Phys. Rev. Lett.}\ }\textbf {\bibinfo
  {volume} {17}},\ \bibinfo {pages} {1133} (\bibinfo {year}
  {1966})}\BibitemShut {NoStop}%
\bibitem [{\citenamefont {{Villain, J.}}(1977)}]{Villain1977}%
  \BibitemOpen
  \bibfield  {author} {\bibinfo {author} {\bibnamefont {{Villain, J.}}},\
  }\href {\doibase 10.1051/jphys:01977003804038500} {\bibfield  {journal}
  {\bibinfo  {journal} {J. Phys. France}\ }\textbf {\bibinfo {volume} {38}},\
  \bibinfo {pages} {385} (\bibinfo {year} {1977})}\BibitemShut {NoStop}%
\bibitem [{\citenamefont {{Villain, J.}}\ \emph {et~al.}(1980)\citenamefont
  {{Villain, J.}}, \citenamefont {{Bidaux, R.}}, \citenamefont {{Carton,
  J.-P.}},\ and\ \citenamefont {{Conte, R.}}}]{Villain1980}%
  \BibitemOpen
  \bibfield  {author} {\bibinfo {author} {\bibnamefont {{Villain, J.}}},
  \bibinfo {author} {\bibnamefont {{Bidaux, R.}}}, \bibinfo {author}
  {\bibnamefont {{Carton, J.-P.}}}, \ and\ \bibinfo {author} {\bibnamefont
  {{Conte, R.}}},\ }\href {\doibase 10.1051/jphys:0198000410110126300}
  {\bibfield  {journal} {\bibinfo  {journal} {J. Phys. France}\ }\textbf
  {\bibinfo {volume} {41}},\ \bibinfo {pages} {1263} (\bibinfo {year}
  {1980})}\BibitemShut {NoStop}%
\bibitem [{\citenamefont {Henley}(1989)}]{Henley1989}%
  \BibitemOpen
  \bibfield  {author} {\bibinfo {author} {\bibfnamefont {C.~L.}\ \bibnamefont
  {Henley}},\ }\href {\doibase 10.1103/PhysRevLett.62.2056} {\bibfield
  {journal} {\bibinfo  {journal} {Phys. Rev. Lett.}\ }\textbf {\bibinfo
  {volume} {62}},\ \bibinfo {pages} {2056} (\bibinfo {year}
  {1989})}\BibitemShut {NoStop}%
\bibitem [{\citenamefont {Chandra}\ \emph {et~al.}(1990)\citenamefont
  {Chandra}, \citenamefont {Coleman},\ and\ \citenamefont {Larkin}}]{CCL1990}%
  \BibitemOpen
  \bibfield  {author} {\bibinfo {author} {\bibfnamefont {P.}~\bibnamefont
  {Chandra}}, \bibinfo {author} {\bibfnamefont {P.}~\bibnamefont {Coleman}}, \
  and\ \bibinfo {author} {\bibfnamefont {A.~I.}\ \bibnamefont {Larkin}},\
  }\href {\doibase 10.1103/PhysRevLett.64.88} {\bibfield  {journal} {\bibinfo
  {journal} {Phys. Rev. Lett.}\ }\textbf {\bibinfo {volume} {64}},\ \bibinfo
  {pages} {88} (\bibinfo {year} {1990})}\BibitemShut {NoStop}%
\bibitem [{\citenamefont {Mulder}\ \emph {et~al.}(2010)\citenamefont {Mulder},
  \citenamefont {Ganesh}, \citenamefont {Capriotti},\ and\ \citenamefont
  {Paramekanti}}]{Mulder2010}%
  \BibitemOpen
  \bibfield  {author} {\bibinfo {author} {\bibfnamefont {A.}~\bibnamefont
  {Mulder}}, \bibinfo {author} {\bibfnamefont {R.}~\bibnamefont {Ganesh}},
  \bibinfo {author} {\bibfnamefont {L.}~\bibnamefont {Capriotti}}, \ and\
  \bibinfo {author} {\bibfnamefont {A.}~\bibnamefont {Paramekanti}},\ }\href
  {\doibase 10.1103/PhysRevB.81.214419} {\bibfield  {journal} {\bibinfo
  {journal} {Phys. Rev. B}\ }\textbf {\bibinfo {volume} {81}},\ \bibinfo
  {pages} {214419} (\bibinfo {year} {2010})}\BibitemShut {NoStop}%
\bibitem [{\citenamefont {Okumura}\ \emph {et~al.}(2010)\citenamefont
  {Okumura}, \citenamefont {Kawamura}, \citenamefont {Okubo},\ and\
  \citenamefont {Motome}}]{Okumura2010}%
  \BibitemOpen
  \bibfield  {author} {\bibinfo {author} {\bibfnamefont {S.}~\bibnamefont
  {Okumura}}, \bibinfo {author} {\bibfnamefont {H.}~\bibnamefont {Kawamura}},
  \bibinfo {author} {\bibfnamefont {T.}~\bibnamefont {Okubo}}, \ and\ \bibinfo
  {author} {\bibfnamefont {Y.}~\bibnamefont {Motome}},\ }\href {\doibase
  10.1143/JPSJ.79.114705} {\bibfield  {journal} {\bibinfo  {journal} {Journal
  of the Physical Society of Japan}\ }\textbf {\bibinfo {volume} {79}},\
  \bibinfo {pages} {114705} (\bibinfo {year} {2010})},\ \Eprint
  {http://arxiv.org/abs/https://doi.org/10.1143/JPSJ.79.114705}
  {https://doi.org/10.1143/JPSJ.79.114705} \BibitemShut {NoStop}%
\bibitem [{\citenamefont {Seabra}\ \emph {et~al.}(2016)\citenamefont {Seabra},
  \citenamefont {Sindzingre}, \citenamefont {Momoi},\ and\ \citenamefont
  {Shannon}}]{Seabra2016}%
  \BibitemOpen
  \bibfield  {author} {\bibinfo {author} {\bibfnamefont {L.}~\bibnamefont
  {Seabra}}, \bibinfo {author} {\bibfnamefont {P.}~\bibnamefont {Sindzingre}},
  \bibinfo {author} {\bibfnamefont {T.}~\bibnamefont {Momoi}}, \ and\ \bibinfo
  {author} {\bibfnamefont {N.}~\bibnamefont {Shannon}},\ }\href {\doibase
  10.1103/PhysRevB.93.085132} {\bibfield  {journal} {\bibinfo  {journal} {Phys.
  Rev. B}\ }\textbf {\bibinfo {volume} {93}},\ \bibinfo {pages} {085132}
  (\bibinfo {year} {2016})}\BibitemShut {NoStop}%
\bibitem [{\citenamefont {Rastelli}\ \emph {et~al.}(1979)\citenamefont
  {Rastelli}, \citenamefont {Tassi},\ and\ \citenamefont
  {Reatto}}]{Rastelli1979}%
  \BibitemOpen
  \bibfield  {author} {\bibinfo {author} {\bibfnamefont {E.}~\bibnamefont
  {Rastelli}}, \bibinfo {author} {\bibfnamefont {A.}~\bibnamefont {Tassi}}, \
  and\ \bibinfo {author} {\bibfnamefont {L.}~\bibnamefont {Reatto}},\ }\href
  {\doibase 10.1016/0378-4363(79)90002-0} {\bibfield  {journal} {\bibinfo
  {journal} {Physica B}\ }\textbf {\bibinfo {volume} {97}},\ \bibinfo {pages}
  {1} (\bibinfo {year} {1979})}\BibitemShut {NoStop}%
\bibitem [{\citenamefont {Okubo}\ \emph {et~al.}(2012)\citenamefont {Okubo},
  \citenamefont {Chung},\ and\ \citenamefont {Kawamura}}]{Okubo2012}%
  \BibitemOpen
  \bibfield  {author} {\bibinfo {author} {\bibfnamefont {T.}~\bibnamefont
  {Okubo}}, \bibinfo {author} {\bibfnamefont {S.}~\bibnamefont {Chung}}, \ and\
  \bibinfo {author} {\bibfnamefont {H.}~\bibnamefont {Kawamura}},\ }\href
  {\doibase 10.1103/PhysRevLett.108.017206} {\bibfield  {journal} {\bibinfo
  {journal} {Phys. Rev. Lett.}\ }\textbf {\bibinfo {volume} {108}},\ \bibinfo
  {pages} {017206} (\bibinfo {year} {2012})}\BibitemShut {NoStop}%
\bibitem [{\citenamefont {Nakatsuji}\ \emph {et~al.}(2007)\citenamefont
  {Nakatsuji}, \citenamefont {Nambu}, \citenamefont {Onuma}, \citenamefont
  {Jonas}, \citenamefont {Broholm},\ and\ \citenamefont
  {Maeno}}]{Nakatsuji2007}%
  \BibitemOpen
  \bibfield  {author} {\bibinfo {author} {\bibfnamefont {S.}~\bibnamefont
  {Nakatsuji}}, \bibinfo {author} {\bibfnamefont {Y.}~\bibnamefont {Nambu}},
  \bibinfo {author} {\bibfnamefont {K.}~\bibnamefont {Onuma}}, \bibinfo
  {author} {\bibfnamefont {S.}~\bibnamefont {Jonas}}, \bibinfo {author}
  {\bibfnamefont {C.}~\bibnamefont {Broholm}}, \ and\ \bibinfo {author}
  {\bibfnamefont {Y.}~\bibnamefont {Maeno}},\ }\href {\doibase
  10.1088/0953-8984/19/14/145232} {\bibfield  {journal} {\bibinfo  {journal}
  {Journal of Physics: Condensed Matter}\ }\textbf {\bibinfo {volume} {19}},\
  \bibinfo {pages} {145232} (\bibinfo {year} {2007})}\BibitemShut {NoStop}%
\bibitem [{\citenamefont {Mazin}(2007)}]{Mazin2007}%
  \BibitemOpen
  \bibfield  {author} {\bibinfo {author} {\bibfnamefont {I.~I.}\ \bibnamefont
  {Mazin}},\ }\href {\doibase 10.1103/PhysRevB.76.140406} {\bibfield  {journal}
  {\bibinfo  {journal} {Phys. Rev. B}\ }\textbf {\bibinfo {volume} {76}},\
  \bibinfo {pages} {140406(R)} (\bibinfo {year} {2007})}\BibitemShut {NoStop}%
\bibitem [{\citenamefont {Tamura}\ and\ \citenamefont
  {Kawashima}(2008)}]{Tamura2008}%
  \BibitemOpen
  \bibfield  {author} {\bibinfo {author} {\bibfnamefont {R.}~\bibnamefont
  {Tamura}}\ and\ \bibinfo {author} {\bibfnamefont {N.}~\bibnamefont
  {Kawashima}},\ }\href {\doibase 10.1143/JPSJ.77.103002} {\bibfield  {journal}
  {\bibinfo  {journal} {Journal of the Physical Society of Japan}\ }\textbf
  {\bibinfo {volume} {77}},\ \bibinfo {pages} {103002} (\bibinfo {year}
  {2008})},\ \Eprint
  {http://arxiv.org/abs/https://doi.org/10.1143/JPSJ.77.103002}
  {https://doi.org/10.1143/JPSJ.77.103002} \BibitemShut {NoStop}%
\bibitem [{\citenamefont {Iaconis}\ \emph {et~al.}(2018)\citenamefont
  {Iaconis}, \citenamefont {Liu}, \citenamefont {Hal{\'a}sz},\ and\
  \citenamefont {Balents}}]{Iaconis2018}%
  \BibitemOpen
  \bibfield  {author} {\bibinfo {author} {\bibfnamefont {J.}~\bibnamefont
  {Iaconis}}, \bibinfo {author} {\bibfnamefont {C.-X.}\ \bibnamefont {Liu}},
  \bibinfo {author} {\bibfnamefont {G.}~\bibnamefont {Hal{\'a}sz}}, \ and\
  \bibinfo {author} {\bibfnamefont {L.}~\bibnamefont {Balents}},\ }\href
  {https://scipost.org/SciPostPhys.4.1.003} {\bibfield  {journal} {\bibinfo
  {journal} {SciPost Physics}\ }\textbf {\bibinfo {volume} {4}},\ \bibinfo
  {pages} {003} (\bibinfo {year} {2018})}\BibitemShut {NoStop}%
\bibitem [{\citenamefont {Gong}\ \emph {et~al.}(2019)\citenamefont {Gong},
  \citenamefont {Zheng}, \citenamefont {Lee}, \citenamefont {Lu},\ and\
  \citenamefont {Sheng}}]{Gong2019}%
  \BibitemOpen
  \bibfield  {author} {\bibinfo {author} {\bibfnamefont {S.-S.}\ \bibnamefont
  {Gong}}, \bibinfo {author} {\bibfnamefont {W.}~\bibnamefont {Zheng}},
  \bibinfo {author} {\bibfnamefont {M.}~\bibnamefont {Lee}}, \bibinfo {author}
  {\bibfnamefont {Y.-M.}\ \bibnamefont {Lu}}, \ and\ \bibinfo {author}
  {\bibfnamefont {D.~N.}\ \bibnamefont {Sheng}},\ }\href {\doibase
  10.1103/PhysRevB.100.241111} {\bibfield  {journal} {\bibinfo  {journal}
  {Phys. Rev. B}\ }\textbf {\bibinfo {volume} {100}},\ \bibinfo {pages}
  {241111(R)} (\bibinfo {year} {2019})}\BibitemShut {NoStop}%
\bibitem [{\citenamefont {Schecter}\ \emph {et~al.}(2017)\citenamefont
  {Schecter}, \citenamefont {Sylju{\aa}sen},\ and\ \citenamefont
  {Paaske}}]{Schecter2017}%
  \BibitemOpen
  \bibfield  {author} {\bibinfo {author} {\bibfnamefont {M.}~\bibnamefont
  {Schecter}}, \bibinfo {author} {\bibfnamefont {O.~F.}\ \bibnamefont
  {Sylju{\aa}sen}}, \ and\ \bibinfo {author} {\bibfnamefont {J.}~\bibnamefont
  {Paaske}},\ }\href {\doibase 10.1103/PhysRevLett.119.157202} {\bibfield
  {journal} {\bibinfo  {journal} {Phys. Rev. Lett.}\ }\textbf {\bibinfo
  {volume} {119}},\ \bibinfo {pages} {157202} (\bibinfo {year}
  {2017})}\BibitemShut {NoStop}%
\bibitem [{\citenamefont {Sylju\aa{}sen}\ \emph {et~al.}(2019)\citenamefont
  {Sylju\aa{}sen}, \citenamefont {Paaske},\ and\ \citenamefont
  {Schecter}}]{Syljuasen2019}%
  \BibitemOpen
  \bibfield  {author} {\bibinfo {author} {\bibfnamefont {O.~F.}\ \bibnamefont
  {Sylju\aa{}sen}}, \bibinfo {author} {\bibfnamefont {J.}~\bibnamefont
  {Paaske}}, \ and\ \bibinfo {author} {\bibfnamefont {M.}~\bibnamefont
  {Schecter}},\ }\href {\doibase 10.1103/PhysRevB.99.174404} {\bibfield
  {journal} {\bibinfo  {journal} {Phys. Rev. B}\ }\textbf {\bibinfo {volume}
  {99}},\ \bibinfo {pages} {174404} (\bibinfo {year} {2019})}\BibitemShut
  {NoStop}%
\bibitem [{\citenamefont {Barci}\ \emph {et~al.}(2013)\citenamefont {Barci},
  \citenamefont {Mendoza-Coto},\ and\ \citenamefont {Stariolo}}]{Barci2013}%
  \BibitemOpen
  \bibfield  {author} {\bibinfo {author} {\bibfnamefont {D.~G.}\ \bibnamefont
  {Barci}}, \bibinfo {author} {\bibfnamefont {A.}~\bibnamefont {Mendoza-Coto}},
  \ and\ \bibinfo {author} {\bibfnamefont {D.~A.}\ \bibnamefont {Stariolo}},\
  }\href {\doibase 10.1103/PhysRevE.88.062140} {\bibfield  {journal} {\bibinfo
  {journal} {Phys. Rev. E}\ }\textbf {\bibinfo {volume} {88}},\ \bibinfo
  {pages} {062140} (\bibinfo {year} {2013})}\BibitemShut {NoStop}%
\bibitem [{\citenamefont {Tamura}\ and\ \citenamefont
  {Kawashima}(2011)}]{Tamura2011}%
  \BibitemOpen
  \bibfield  {author} {\bibinfo {author} {\bibfnamefont {R.}~\bibnamefont
  {Tamura}}\ and\ \bibinfo {author} {\bibfnamefont {N.}~\bibnamefont
  {Kawashima}},\ }\href {\doibase 10.1143/JPSJ.80.074008} {\bibfield  {journal}
  {\bibinfo  {journal} {Journal of the Physical Society of Japan}\ }\textbf
  {\bibinfo {volume} {80}},\ \bibinfo {pages} {074008} (\bibinfo {year}
  {2011})},\ \Eprint
  {http://arxiv.org/abs/https://doi.org/10.1143/JPSJ.80.074008}
  {https://doi.org/10.1143/JPSJ.80.074008} \BibitemShut {NoStop}%
\bibitem [{\citenamefont {Halperin}\ and\ \citenamefont
  {Nelson}(1978)}]{HalperinNelson78}%
  \BibitemOpen
  \bibfield  {author} {\bibinfo {author} {\bibfnamefont {B.~I.}\ \bibnamefont
  {Halperin}}\ and\ \bibinfo {author} {\bibfnamefont {D.~R.}\ \bibnamefont
  {Nelson}},\ }\href {\doibase 10.1103/PhysRevLett.41.121} {\bibfield
  {journal} {\bibinfo  {journal} {Phys. Rev. Lett.}\ }\textbf {\bibinfo
  {volume} {41}},\ \bibinfo {pages} {121} (\bibinfo {year} {1978})}\BibitemShut
  {NoStop}%
\bibitem [{\citenamefont {Nelson}\ and\ \citenamefont
  {Halperin}(1979)}]{NelsonHalperin79}%
  \BibitemOpen
  \bibfield  {author} {\bibinfo {author} {\bibfnamefont {D.~R.}\ \bibnamefont
  {Nelson}}\ and\ \bibinfo {author} {\bibfnamefont {B.~I.}\ \bibnamefont
  {Halperin}},\ }\href {\doibase 10.1103/PhysRevB.19.2457} {\bibfield
  {journal} {\bibinfo  {journal} {Phys. Rev. B}\ }\textbf {\bibinfo {volume}
  {19}},\ \bibinfo {pages} {2457} (\bibinfo {year} {1979})}\BibitemShut
  {NoStop}%
\bibitem [{\citenamefont {Li}\ \emph {et~al.}(2012)\citenamefont {Li},
  \citenamefont {Nattermann},\ and\ \citenamefont {Pokrovsky}}]{Li2012}%
  \BibitemOpen
  \bibfield  {author} {\bibinfo {author} {\bibfnamefont {F.}~\bibnamefont
  {Li}}, \bibinfo {author} {\bibfnamefont {T.}~\bibnamefont {Nattermann}}, \
  and\ \bibinfo {author} {\bibfnamefont {V.~L.}\ \bibnamefont {Pokrovsky}},\
  }\href {\doibase 10.1103/PhysRevLett.108.107203} {\bibfield  {journal}
  {\bibinfo  {journal} {Phys. Rev. Lett.}\ }\textbf {\bibinfo {volume} {108}},\
  \bibinfo {pages} {107203} (\bibinfo {year} {2012})}\BibitemShut {NoStop}%
\bibitem [{\citenamefont {Schoenherr}\ \emph {et~al.}(2018)\citenamefont
  {Schoenherr}, \citenamefont {M{\"u}ller}, \citenamefont {K{\"o}hler},
  \citenamefont {Rosch}, \citenamefont {Kanazawa}, \citenamefont {Tokura},
  \citenamefont {Garst},\ and\ \citenamefont {Meier}}]{Schoenherr2018}%
  \BibitemOpen
  \bibfield  {author} {\bibinfo {author} {\bibfnamefont {P.}~\bibnamefont
  {Schoenherr}}, \bibinfo {author} {\bibfnamefont {J.}~\bibnamefont
  {M{\"u}ller}}, \bibinfo {author} {\bibfnamefont {L.}~\bibnamefont
  {K{\"o}hler}}, \bibinfo {author} {\bibfnamefont {A.}~\bibnamefont {Rosch}},
  \bibinfo {author} {\bibfnamefont {N.}~\bibnamefont {Kanazawa}}, \bibinfo
  {author} {\bibfnamefont {Y.}~\bibnamefont {Tokura}}, \bibinfo {author}
  {\bibfnamefont {M.}~\bibnamefont {Garst}}, \ and\ \bibinfo {author}
  {\bibfnamefont {D.}~\bibnamefont {Meier}},\ }\href {\doibase
  10.1038/s41567-018-0056-5} {\bibfield  {journal} {\bibinfo  {journal} {Nature
  Physics}\ }\textbf {\bibinfo {volume} {14}},\ \bibinfo {pages} {465}
  (\bibinfo {year} {2018})}\BibitemShut {NoStop}%
\bibitem [{\citenamefont {Nattermann}\ and\ \citenamefont
  {Pokrovsky}(2018)}]{Nattermann2018}%
  \BibitemOpen
  \bibfield  {author} {\bibinfo {author} {\bibfnamefont {T.}~\bibnamefont
  {Nattermann}}\ and\ \bibinfo {author} {\bibfnamefont {V.~L.}\ \bibnamefont
  {Pokrovsky}},\ }\href {\doibase 10.1134/S106377611811016X} {\bibfield
  {journal} {\bibinfo  {journal} {Journal of Experimental and Theoretical
  Physics}\ }\textbf {\bibinfo {volume} {127}},\ \bibinfo {pages} {922}
  (\bibinfo {year} {2018})}\BibitemShut {NoStop}%
\bibitem [{\citenamefont {Korshunov}(2005)}]{Korshunov2005}%
  \BibitemOpen
  \bibfield  {author} {\bibinfo {author} {\bibfnamefont {S.~E.}\ \bibnamefont
  {Korshunov}},\ }\href {\doibase 10.1103/PhysRevB.72.144417} {\bibfield
  {journal} {\bibinfo  {journal} {Phys. Rev. B}\ }\textbf {\bibinfo {volume}
  {72}},\ \bibinfo {pages} {144417} (\bibinfo {year} {2005})}\BibitemShut
  {NoStop}%
\bibitem [{\citenamefont {Smerald}\ \emph {et~al.}(2016)\citenamefont
  {Smerald}, \citenamefont {Korshunov},\ and\ \citenamefont
  {Mila}}]{Smerald2016}%
  \BibitemOpen
  \bibfield  {author} {\bibinfo {author} {\bibfnamefont {A.}~\bibnamefont
  {Smerald}}, \bibinfo {author} {\bibfnamefont {S.}~\bibnamefont {Korshunov}},
  \ and\ \bibinfo {author} {\bibfnamefont {F.}~\bibnamefont {Mila}},\ }\href
  {\doibase 10.1103/PhysRevLett.116.197201} {\bibfield  {journal} {\bibinfo
  {journal} {Phys. Rev. Lett.}\ }\textbf {\bibinfo {volume} {116}},\ \bibinfo
  {pages} {197201} (\bibinfo {year} {2016})}\BibitemShut {NoStop}%
\bibitem [{\citenamefont {Kawamura}\ and\ \citenamefont
  {Miyashita}(1984)}]{KawamuraMiyashita1984}%
  \BibitemOpen
  \bibfield  {author} {\bibinfo {author} {\bibfnamefont {H.}~\bibnamefont
  {Kawamura}}\ and\ \bibinfo {author} {\bibfnamefont {S.}~\bibnamefont
  {Miyashita}},\ }\href {\doibase 10.1143/JPSJ.53.4138} {\bibfield  {journal}
  {\bibinfo  {journal} {Journal of the Physical Society of Japan}\ }\textbf
  {\bibinfo {volume} {53}},\ \bibinfo {pages} {4138} (\bibinfo {year}
  {1984})},\ \Eprint
  {http://arxiv.org/abs/https://doi.org/10.1143/JPSJ.53.4138}
  {https://doi.org/10.1143/JPSJ.53.4138} \BibitemShut {NoStop}%
\bibitem [{\citenamefont {Robert}\ \emph {et~al.}(2008)\citenamefont {Robert},
  \citenamefont {Canals}, \citenamefont {Simonet},\ and\ \citenamefont
  {Ballou}}]{Robert2008}%
  \BibitemOpen
  \bibfield  {author} {\bibinfo {author} {\bibfnamefont {J.}~\bibnamefont
  {Robert}}, \bibinfo {author} {\bibfnamefont {B.}~\bibnamefont {Canals}},
  \bibinfo {author} {\bibfnamefont {V.}~\bibnamefont {Simonet}}, \ and\
  \bibinfo {author} {\bibfnamefont {R.}~\bibnamefont {Ballou}},\ }\href
  {\doibase 10.1103/PhysRevLett.101.117207} {\bibfield  {journal} {\bibinfo
  {journal} {Phys. Rev. Lett.}\ }\textbf {\bibinfo {volume} {101}},\ \bibinfo
  {pages} {117207} (\bibinfo {year} {2008})}\BibitemShut {NoStop}%
\bibitem [{\citenamefont {Guitteny}\ \emph {et~al.}(2013)\citenamefont
  {Guitteny}, \citenamefont {Robert}, \citenamefont {Bonville}, \citenamefont
  {Ollivier}, \citenamefont {Decorse}, \citenamefont {Steffens}, \citenamefont
  {Boehm}, \citenamefont {Mutka}, \citenamefont {Mirebeau},\ and\ \citenamefont
  {Petit}}]{Guitteny2013}%
  \BibitemOpen
  \bibfield  {author} {\bibinfo {author} {\bibfnamefont {S.}~\bibnamefont
  {Guitteny}}, \bibinfo {author} {\bibfnamefont {J.}~\bibnamefont {Robert}},
  \bibinfo {author} {\bibfnamefont {P.}~\bibnamefont {Bonville}}, \bibinfo
  {author} {\bibfnamefont {J.}~\bibnamefont {Ollivier}}, \bibinfo {author}
  {\bibfnamefont {C.}~\bibnamefont {Decorse}}, \bibinfo {author} {\bibfnamefont
  {P.}~\bibnamefont {Steffens}}, \bibinfo {author} {\bibfnamefont
  {M.}~\bibnamefont {Boehm}}, \bibinfo {author} {\bibfnamefont
  {H.}~\bibnamefont {Mutka}}, \bibinfo {author} {\bibfnamefont
  {I.}~\bibnamefont {Mirebeau}}, \ and\ \bibinfo {author} {\bibfnamefont
  {S.}~\bibnamefont {Petit}},\ }\href {\doibase 10.1103/PhysRevLett.111.087201}
  {\bibfield  {journal} {\bibinfo  {journal} {Phys. Rev. Lett.}\ }\textbf
  {\bibinfo {volume} {111}},\ \bibinfo {pages} {087201} (\bibinfo {year}
  {2013})}\BibitemShut {NoStop}%
\bibitem [{\citenamefont {Yan}\ \emph {et~al.}(2018)\citenamefont {Yan},
  \citenamefont {Pohle},\ and\ \citenamefont {Shannon}}]{Yan2018}%
  \BibitemOpen
  \bibfield  {author} {\bibinfo {author} {\bibfnamefont {H.}~\bibnamefont
  {Yan}}, \bibinfo {author} {\bibfnamefont {R.}~\bibnamefont {Pohle}}, \ and\
  \bibinfo {author} {\bibfnamefont {N.}~\bibnamefont {Shannon}},\ }\href
  {\doibase 10.1103/PhysRevB.98.140402} {\bibfield  {journal} {\bibinfo
  {journal} {Phys. Rev. B}\ }\textbf {\bibinfo {volume} {98}},\ \bibinfo
  {pages} {140402(R)} (\bibinfo {year} {2018})}\BibitemShut {NoStop}%
\bibitem [{\citenamefont {Mizoguchi}\ \emph {et~al.}(2018)\citenamefont
  {Mizoguchi}, \citenamefont {Jaubert}, \citenamefont {Moessner},\ and\
  \citenamefont {Udagawa}}]{Mizoguchi2018}%
  \BibitemOpen
  \bibfield  {author} {\bibinfo {author} {\bibfnamefont {T.}~\bibnamefont
  {Mizoguchi}}, \bibinfo {author} {\bibfnamefont {L.~D.~C.}\ \bibnamefont
  {Jaubert}}, \bibinfo {author} {\bibfnamefont {R.}~\bibnamefont {Moessner}}, \
  and\ \bibinfo {author} {\bibfnamefont {M.}~\bibnamefont {Udagawa}},\ }\href
  {\doibase 10.1103/PhysRevB.98.144446} {\bibfield  {journal} {\bibinfo
  {journal} {Phys. Rev. B}\ }\textbf {\bibinfo {volume} {98}},\ \bibinfo
  {pages} {144446} (\bibinfo {year} {2018})}\BibitemShut {NoStop}%
\bibitem [{\citenamefont {Fang}\ \emph {et~al.}(2008)\citenamefont {Fang},
  \citenamefont {Yao}, \citenamefont {Tsai}, \citenamefont {Hu},\ and\
  \citenamefont {Kivelson}}]{Fang2008}%
  \BibitemOpen
  \bibfield  {author} {\bibinfo {author} {\bibfnamefont {C.}~\bibnamefont
  {Fang}}, \bibinfo {author} {\bibfnamefont {H.}~\bibnamefont {Yao}}, \bibinfo
  {author} {\bibfnamefont {W.-F.}\ \bibnamefont {Tsai}}, \bibinfo {author}
  {\bibfnamefont {J.~P.}\ \bibnamefont {Hu}}, \ and\ \bibinfo {author}
  {\bibfnamefont {S.~A.}\ \bibnamefont {Kivelson}},\ }\href {\doibase
  10.1103/PhysRevB.77.224509} {\bibfield  {journal} {\bibinfo  {journal} {Phys.
  Rev. B}\ }\textbf {\bibinfo {volume} {77}},\ \bibinfo {pages} {224509}
  (\bibinfo {year} {2008})}\BibitemShut {NoStop}%
\bibitem [{\citenamefont {Jepsen}\ \emph {et~al.}(2020)\citenamefont {Jepsen},
  \citenamefont {Amato-Grill}, \citenamefont {Dimitrova}, \citenamefont {Ho},
  \citenamefont {Demler},\ and\ \citenamefont {Ketterle}}]{Jepsen2020}%
  \BibitemOpen
  \bibfield  {author} {\bibinfo {author} {\bibfnamefont {P.~N.}\ \bibnamefont
  {Jepsen}}, \bibinfo {author} {\bibfnamefont {J.}~\bibnamefont {Amato-Grill}},
  \bibinfo {author} {\bibfnamefont {I.}~\bibnamefont {Dimitrova}}, \bibinfo
  {author} {\bibfnamefont {W.~W.}\ \bibnamefont {Ho}}, \bibinfo {author}
  {\bibfnamefont {E.}~\bibnamefont {Demler}}, \ and\ \bibinfo {author}
  {\bibfnamefont {W.}~\bibnamefont {Ketterle}},\ }\href {\doibase
  10.1038/s41586-020-3033-y} {\bibfield  {journal} {\bibinfo  {journal}
  {Nature}\ }\textbf {\bibinfo {volume} {588}},\ \bibinfo {pages} {403}
  (\bibinfo {year} {2020})}\BibitemShut {NoStop}%
\bibitem [{\citenamefont {Bergman}\ \emph {et~al.}(2007)\citenamefont
  {Bergman}, \citenamefont {Alicea}, \citenamefont {Gull}, \citenamefont
  {Trebst},\ and\ \citenamefont {Balents}}]{Bergman2007}%
  \BibitemOpen
  \bibfield  {author} {\bibinfo {author} {\bibfnamefont {D.}~\bibnamefont
  {Bergman}}, \bibinfo {author} {\bibfnamefont {J.}~\bibnamefont {Alicea}},
  \bibinfo {author} {\bibfnamefont {E.}~\bibnamefont {Gull}}, \bibinfo {author}
  {\bibfnamefont {S.}~\bibnamefont {Trebst}}, \ and\ \bibinfo {author}
  {\bibfnamefont {L.}~\bibnamefont {Balents}},\ }\href {\doibase
  10.1038/nphys622} {\bibfield  {journal} {\bibinfo  {journal} {Nature
  Physics}\ }\textbf {\bibinfo {volume} {3}},\ \bibinfo {pages} {487} (\bibinfo
  {year} {2007})}\BibitemShut {NoStop}%
\end{thebibliography}%

\end{document}